\documentclass[aps,pre,preprint,groupedaddress,showpacs]{revtex4-1}
\usepackage{amssymb}

\usepackage[dvips]{graphicx}
\usepackage{textcomp}
\usepackage{amssymb}

\begin{document}

\title{
\Large\bf Thermodynamic Casimir Forces between a Sphere and a Plate: 
          Monte Carlo Simulation of a Spin Model}

\author{Martin Hasenbusch}
\email[]{Martin.Hasenbusch@physik.hu-berlin.de}
\affiliation{
Institut f\"ur Physik, Humboldt-Universit\"at zu Berlin,
Newtonstr. 15, 12489 Berlin, Germany}

\date{\today}

\begin{abstract}
We study the thermodynamic Casimir force between a spherical object
and a plate. We consider the bulk universality class of the three-dimensional
Ising model, which is relevant for experiments on binary mixtures. 
To this end, we simulate the improved Blume-Capel model.
Following Hucht, we compute the force by integrating 
energy differences over the inverse temperature. We demonstrate that these 
energy differences can be computed efficiently by using a particular cluster algorithm.
Our numerical results for strongly symmetry breaking boundary conditions 
are compared with the Derjaguin approximation for small 
distances and the small sphere expansion for large distances between the sphere 
and the plate. 
\end{abstract}

\maketitle

\section{Introduction}
In 1978  Fisher and de Gennes \cite{FiGe78} realized that when thermal
fluctuations are restricted by the finite extension of the system, a force acts on 
the boundaries. Since this effect is  similar to the Casimir effect,
where the restriction of quantum fluctuations induces a force, it is called
thermodynamic Casimir effect. Recently this force could be
detected for various experimental systems and quantitative predictions could
be obtained from Monte Carlo simulations of spin models \cite{Ga09}.
The thermodynamic Casimir force is experimentally and maybe technologically interesting
since it depends, in contrast to other forces, strongly on the temperature.
Hence, since the temperature can be easily changed in experiments,
the thermodynamic Casimir force can be easily manipulated.
For a discussion of the thermodynamic Casimir force in the general context of nanoscale 
science see ref. \cite{Frenchetal}.

Experiments on thin films of $^4$He and mixtures of $^3$He and $^4$He near the 
$\lambda$-transition  \cite{GaCh99,GaCh02,GaScGaCh06} and 
on thin films of binary mixtures near the mixing-demixing transition \cite{FuYaPe05,RaBoJa07} 
have been performed. Theoretically
these experiments are well described by a plate-plate geometry. For this geometry
the thermodynamic Casimir force per area $F_{C}$ follows the finite size scaling form
\cite{Krech}
\begin{equation}
\label{Tplateplate}
 F_{C}  \simeq k_B T  L^{-d}  \; \theta(t [L/\xi_{0,+}]^{1/\nu}) \;\;,
\end{equation}
where $L$ is the thickness of the film, $t$ is the reduced temperature,
$d$ is the dimension of the bulk system and $\theta$ is a universal
function that depends on the bulk and boundary universality classes that characterize
the film. The correlation length in the high temperature phase behaves as 
$\xi \simeq \xi_{0,+} t^{-\nu}$, where $\xi_{0,+}$ is the amplitude of 
the correlation length in the high temperature phase and $\nu$ is the critical 
exponent of the correlation length. 

Field theoretic methods do not allow to compute $\theta$ for three-dimensional 
systems in the full range of the scaling variable \cite{Diehl86,Diehl97}. Therefore 
the fact that $\theta$ recently has been computed quite accurately by using Monte Carlo
simulations of spin models constitutes valuable progress. In refs. \cite{Hucht,VaGaMaDi08,Ha09,Ha10}
the three-dimensional XY bulk universality class and a vanishing field at the boundary have been 
studied, which is relevant for the experiments on $^4$He. A quite satisfactory agreement 
between the experimental results and the theory was found. In refs.
\cite{DaKr04,VaGaMaDi07,VaGaMaDi08,mybreaking,PaToDi10,mycrossover,VaMaDi11,HuGrSc11,mycorrection}
the Ising bulk universality class and various types of boundary conditions were studied. 
Note that the mixing-demixing transition of binary mixtures belongs to the Ising bulk 
universality class.

Experimentally it is easier to access the thermodynamic Casimir force for
a sphere-sphere or a sphere-plate than for the plate-plate geometry.
Several experiments for these geometries have been performed.  In particular, the 
thermodynamic Casimir force between a colloidal particle immersed in a binary mixture 
of fluids and the substrate has been directly measured \cite{Nature08,Gaetal09}.
In ref. \cite{NeHeBe09} it was demonstrated that the thermodynamic Casimir forces in 
a colloidal system can be continuously tuned by the choice of the boundary conditions.
In addition to homogeneous substrates also chemically patterned ones have been
studied \cite{So08,Tr11,Zvyagolskaya}.
In other experiments, the thermodynamic Casimir force is the driving force
for colloidal aggregation, see e.g. refs. \cite{Bonnetal09,ZvArBe11}.

For simplicity, we restrict the following discussion to the sphere-plate geometry. 
Furthermore we assume that both the surfaces of the sphere and the plate are
homogeneous. In this case, the thermodynamic Casimir force follows the 
scaling form \cite{HaScEiDi98} 
\begin{equation}
\label{Tsphere}
 F_C(D,R,T) = \frac{k_B T}{R} K(t [D/\xi_{0,+}]^{1/\nu},\Delta)
\end{equation}
where $R$ is the radius of the sphere, $D$ the distance between the plate and the sphere and $\Delta = D/R$.
Note that in the literature often $D/\xi$, where $\xi$ is the bulk correlation length in the high 
temperature phase, is used instead of $t [D/\xi_{0,+}]^{1/\nu}$ as argument of the universal scaling function.
The thermodynamic Casimir force is given by
\begin{equation}
\label{DefineFC}
 F_C(D,R,T) = -  k_B T  \; \frac{ \partial F(D,R,T) }{ \partial D} \;\;
\end{equation}
where $F(D,R,T)$ is the reduced free energy of the system.

In the limit $D \ll R$, the function $K(x,\Delta)$ can be obtained from 
$\theta(t [L/\xi_{0,+}]^{1/\nu})$ by using the Derjaguin approximation (DA) \cite{Derja}.
Instead, in the limit $D \gg R$  the small sphere expansion \cite{BuEi97} can be used.
In ref. \cite{HaScEiDi98} the function $K$ is discussed for the case where both surfaces
strongly adsorb the same component of the binary mixture. The behaviour of $K$ for 
three-dimensional systems for $D \simeq R$ is inferred  from the exactly known one 
in two dimensions and from mean-field calculations.  In ref. \cite{troendle} the thermodynamic
Casimir force is discussed for colloids close to  chemically patterned substrates. To this end the authors
employ the Derjaguin approximation. They check the range of applicability of the Derjaguin 
approximation by using mean-field calculations. In section III of \cite{troendle}, they  
discuss for a start also homogeneous substrates.

Here we discuss a method to determine the thermodynamic Casimir force for the  sphere-plate 
geometry by using Monte Carlo simulations of spin models. The  generalization to the 
sphere-sphere geometry is straightforward. The method is based on the geometric cluster algorithm of 
Heringa and Bl\"ote \cite{HeBl98}.
In the present work, we simulate the improved Blume-Capel model on the simple cubic lattice.
We focus on strongly symmetry breaking boundary conditions at the surfaces of the 
sphere and the plate. We assume that the surface of the plate prefers positive 
magnetisation. We study the two cases $s_{sphere}=-1$ and $1$, where $s_{sphere}$ 
is the value of the spins at the surface of the sphere. Preliminary results for $s_{sphere}=0$,
indicate that also this case can be efficiently simulated. 
In our numerical tests we studied spheres of the radii $R=3.5, 4.5, 7.5$ and $15.5$
lattice spacings.  For $R=3.5$ and $4.5$ we considered distances between the sphere and the 
plate up to $D \approx 10 R$.  We introduce an effective radius $R_{eff}$ of the sphere which 
is determined by a particular finite size scaling method. To this end we compute the ratio of partition 
functions $Z_-/Z_+$ by using a cluster algorithm which is closely related to the 
boundary flip algorithm \cite{MH93A,MH93B}. The index of $Z$ gives the sign of $s_{sphere}$. 
In this finite size scaling study we consider $\Delta \approx 25$, which is large compared with 
the values used otherwise. Analysing our data we find that replacing $R$ by $R_{eff}$  leads
to a good scaling behaviour for the whole range of $\Delta$ studied here.  As result we obtain 
reliable estimates for the scaling functions $K_-(x,\Delta)$ and $K_+(x,\Delta)$ for 
$1 \lessapprox \Delta \lessapprox 12$, where the subscript of $K$ gives the sign of $s_{sphere}$.
The most striking observation is that in the low temperature phase 
$K_-(x,\Delta)$ deviates quite strongly from the Derjaguin approximation already for 
$\Delta \approx 1$ and likely smaller. 

The present work should be seen as pilot study that allows for a number of improvements
as discussed in section \ref{conclus}. The method discussed here should apply to a 
broader range of problems as we also shall discuss in section \ref{conclus}.

The paper is organised as follows. First we define the model. Then we discuss the geometry of 
the systems that we study. In section \ref{algorithm} we discuss the algorithm that is used 
to compute the thermodynamic Casimir force. In section \ref{numerical} we present our numerical
results. First we discuss the performance of the algorithm. Then we give our results for the
scaling functions $K_\pm(x,\Delta)$. In section \ref{conclus} we conclude and give an outlook. 
In Appendix \ref{App1} we discuss the finite size scaling method that we 
used to determine the effective radius $R_{eff}$.

\section{The Model}
We  simulate the Blume-Capel model on the simple cubic 
lattice. It is defined by the reduced Hamiltonian
\begin{equation}
\label{Isingaction}
H = -\beta \sum_{<xy>}  s_x s_y
  + \tilde D \sum_x s_x^2 - \tilde h \sum_x s_x \;\; ,
\end{equation}
where $s_x \in \{-1, 0, 1 \}$ and $x=(x_0,x_1,x_2)$ 
denotes a site on the simple cubic lattice, where $x_i$  takes integer values
and $<xy>$ is a pair of nearest neighbours on the lattice. The partition function
is given by $Z = \sum_{\{s\}} \exp(- H)$, where the sum runs over all spin
configurations. The parameter $\tilde D$ controls the
density of vacancies $s_x=0$. In the limit $\tilde D \rightarrow - \infty$
vacancies are completely suppressed and hence the spin-1/2 Ising
model is recovered. Here we consider a vanishing external field $\tilde h=0$ throughout.
In  $d\ge 2$  dimensions the model undergoes a continuous phase transition
for $-\infty \le  \tilde D   < \tilde D_{tri} $ at a $\beta_c$ that depends on $ \tilde D$.
For $\tilde D > \tilde D_{tri}$ the model undergoes a first order phase transition, where
$\tilde D_{tri}=2.0313(4)$ \cite{DeBl04}. 

Numerically, using Monte Carlo simulations it has been shown that there
is a point $(\tilde D^*,\beta_c(\tilde D^*))$
on the line of second order phase transitions, where the amplitude
of leading corrections to scaling vanishes.  Our recent
estimate is $\tilde D^*=0.656(20)$ \cite{mycritical}.  In \cite{mycritical} we
simulated the model at $\tilde D=0.655$ close to $\beta_c$ on lattices of a
linear size up to $L=360$. From a standard finite size scaling analysis
of phenomenological couplings like the Binder cumulant we find
$\beta_c(0.655)=0.387721735(25)$. Furthermore the amplitude of leading
corrections to scaling is at least by a factor of $30$ smaller than
for the spin-1/2 Ising model. In order to set the scale we use the second 
moment correlation length in the high temperature phase. Its amplitude is 
given by \cite{mycorrection}
\begin{eqnarray}
\label{xi0}
\xi_{2nd,0,+} &=&  0.2283(1) - 1.8 \times (\nu-0.63002)
                        + 275 \times (\beta_c - 0.387721735) \;\; \nonumber \\
&&  \mbox{using} \;\; t = \beta_c - \beta \;\;
 \mbox{as definition of the reduced temperature} .
\end{eqnarray}
In the high temperature phase there is little difference between
$\xi_{2nd}$ and the exponential correlation length $\xi_{exp}$ which
is defined by the asymptotic decay of the two-point correlation function.
Following  \cite{pisaseries}:
\begin{equation}
\lim_{t\searrow 0} \frac{\xi_{exp}}{\xi_{2nd}} = 1.000200(3)
\;\;
\end{equation}
for the thermodynamic limit of the three-dimensional system.
Note that in the following $\xi_{0}$ always refers to $\xi_{2nd,0,+}$.

\section{Geometry of the System}
We have used fixed boundary conditions in $0$-direction and periodic 
boundary conditions in $1$ and $2$ directions. In most of our simulations we have fixed 
$s_x=1$ for $x_0=0$ and $x_0=L_0+1$. In order to check the effect of the second
boundary, we have performed some simulations with $s_x=1$ for $x_0=0$ and 
$s_x=0$ for $x_0=L_0+1$. In both cases there are $L_0$ layers
with fluctuating spins. For $i=1,2$ we take $x_i=-L_i/2+1$,$-L_i/2+2$,$...$,
$L_i/2$ and $L_1=L_2=L$ throughout. 

To model the spherical object, we have fixed all spins $s_x$ to $s_{sphere}$ for
\begin{equation}
\sqrt{(x_0 - h)^2 + x_1^2 + x_2^2} \le R
\end{equation}
The distance between the sphere and the plate is given by 
\begin{equation}
 D = h - R \;\;.
\end{equation}
Analysing our data we shall introduce an effective radius $R_{eff}$  of 
the sphere that we determine by using a finite size scaling analysis in Appendix \ref{App1}
below. Taking also into account the extrapolation length $l_{eff}$ at 
the boundary of the plate we arrive at
\begin{equation}
\label{Deff}
 D_{eff} = h - R_{eff} + l_{eff}
\end{equation}
where $l_{eff} =0.96(2)-0.5=0.46(2)$ for the model and boundary conditions discussed here
\cite{mycrossover}. Note that $l_{eff} =0.96(2)$ given in ref. \cite{mycrossover}  refers
to $x_0=0.5$ as location of the boundary. 
Mostly we have simulated the two cases $s_{sphere}=-1$ and $1$. In our study, 
nearest neighbour pairs that contain a fixed and a fluctuating spin are coupled
by the same $\beta$ as nearest neighbour pairs that contain two fluctuating spins.
One could imagine to chose the couplings 
depending on the position on the surface of the sphere, in order to 
effectively improve the spherical shape of the object. 

\section{The algorithm}
\label{algorithm}
In order to determine the thermodynamic Casimir force $F_C(D,R,\beta)$, eq.~(\ref{DefineFC}),  
between the plate and the 
sphere, we compute finite differences of the reduced free energy $F(D,R,\beta)$ of the system
\begin{equation}
-\beta F_C(D,R,\beta) \approx \Delta F(D,R,\beta) = F(D+1/2,R,\beta)-F(D-1/2,R,\beta) \;\;.
\end{equation}
Note that we chose the  difference of the distances equal to one due to
the discrete translational 
invariance of the lattice. Similarly to the proposal of Hucht \cite{Hucht}, we determine 
$\Delta F(D,R,\beta)$ by integrating the corresponding difference of energies
\begin{equation}
\label{fundamental}
\Delta F(D,R,\beta)= - \int_{\beta_0}^{\beta} \mbox{d} \tilde \beta \; 
\Delta E(D,R,\tilde \beta) 
\end{equation}
where $\Delta E(D,R,\beta) = E(D+1/2,R,\beta)-E(D-1/2,R,\beta)$ and
\begin{equation}
\label{enesummation}
 E(D,R,\beta) = \left \langle \sum_{<xy>} s_x s_y \right \rangle_{D,R,\beta} \;\;.
\end{equation}
The value of $\beta_0$ is chosen such that $\Delta E(D,R,\beta_0) \approx 0$, which is the 
case for $D \gg \xi_{bulk}(\beta_0)$.  The integration is done numerically, using  
the trapezoidal rule. It turns out that similar to the study of the plate-plate 
geometry $O(100)$ nodes are needed to compute the thermodynamic Casimir force in the whole 
range of temperatures that is of interest to us.

\begin{figure}
\begin{center}
\includegraphics[width=8.0cm]{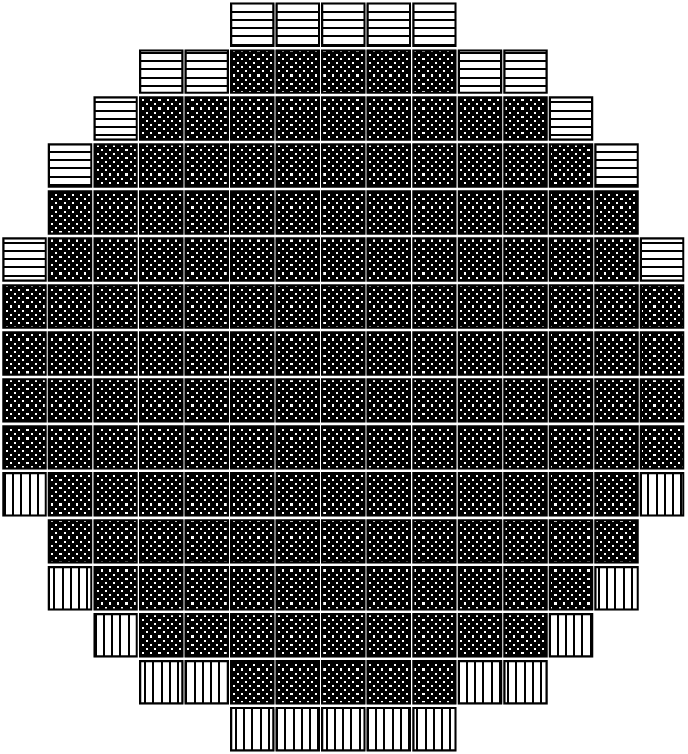}
\caption{\label{ball_sketch}
Two-dimensional sketch of two spheres at $x_0 = D-1/2$ and $D+1/2$. The $0$-axis 
is drawn in vertical direction. The overlapp of the two spheres is given by black 
squares with white dots. Sites that only belong to the lower sphere are represented
by squares with vertical lines and those that only belong to the upper sphere
by squares with horizontal lines.
}
\end{center}
\end{figure}

Routinely one would carry out this programme by simulating the systems
for $D-1/2$ and $D+1/2$ separately using standard Monte Carlo algorithms. In order to
avoid systematic errors,  $L_0,L \gg D,R$ should be taken.
However the variance of $E(D,R,\beta)$ and the CPU-time required for the simulations 
are increasing with the system size. In fact, first numerical experiments
have shown that even for moderate values of $D$ and $R$, this approach fails 
to give sufficiently accurate results.

The expectation value $<s_x s_y>$ only differs strongly between the two systems 
for nearest neighbour pairs $<xy>$ close to the sphere and this
difference decays at least power-like 
with increasing distance from the sphere. Therefore, computing $\Delta E(D,R,\beta)$ one 
would like to concentrate on the part of the system, where the difference in $<s_x s_y>$ 
is large and to avoid to pick up variance from the remainder. Routinely one 
could implement 
this idea by ad hoc restricting the range of summation in eq.~(\ref{enesummation}).  
The problem is that this ad hoc restriction leads to a systematic error that needs 
to be controlled, e.g. by comparing the results of different restrictions.
In the following we shall discuss an algorithm that generates the restriction of the 
summation to the neighbourhood of the sphere automatically and at the same time  
allows to compute $\Delta E(D,R,\beta)$ without bias. 

This algorithm is closely related to the geometric cluster algorithm \cite{HeBl98}.
Similar to ref. \cite{SwWa86}, two systems are updated jointly. These two 
systems share the values of $\beta,L_0,L$ and $R$. The distance of the sphere from the 
boundary is $D-1/2$ and $D+1/2$ for system 1 and 2, respectively. In the following 
the spins are denoted by  $s_{x,i}$, where the second index indicates whether the spin 
belongs to system 1 or 2. Configurations are denoted by $\{s\}_i$. 
The combined partition function is given by
\begin{equation}
 Z_{com} = \sum_{\{s\}_1} \sum_{\{s\}_2} \exp\left[-H_1(\{s\}_1)-H_2(\{s\}_2)\right]
\end{equation}
where $H_i$  is the reduced Hamiltonian of the system $i$. 
The cluster update  comprises the exchange of spins
\begin{eqnarray}
 s_{x,1}' &=&  s_{x,2} \nonumber \\
 s_{x,2}' &=&  s_{x,1}
\end{eqnarray}
for all sites $x$ in the cluster. Performing this exchange of spins for all sites
within a cluster is called flipping the cluster in the following.
A cluster is build in the usual way: A pair of nearest neighbours $<xy>$
is deleted with the probability $p_d$ and frozen otherwise. In our case \cite{HeBl98}
\begin{equation}
 p_d = \mbox{min}\left[1,\exp\left(-\beta [s_{x,1}-s_{x,2}] [s_{y,1} - s_{y,2}]\right)\right]
\;\;.
\end{equation}
Two sites of the lattice belong to the same cluster if they are connected at least
by one chain of frozen pairs of nearest neighbours.
Various choices of the selection of the clusters to be updated are discussed 
in the literature. In the case of the Swendsen-Wang \cite{SwWa87} algorithm all 
clusters are constructed and then all spins within a given cluster are flipped with 
probability $1/2$. In the single cluster algorithm \cite{Wolff} one first selects randomly
one site of the lattice. Then one constructs only the cluster that contains this site.
All spins within this cluster are flipped. Further choices have been discussed
in the literature \cite{Kerler,wall}. 

In our case, the choice of the clusters to be flipped is dictated by the requirement
that the fixed spins have to keep their value under the update. The exchange of spins
should not be allowed for those sites $x$, where $s_{x,1}$ is fixed while $s_{x,2}$ is not, 
or vice versa. In the following these sites are called frozen sites. In Fig. \ref{ball_sketch}
these sites are represented by squares with vertical or horizontal lines on them.
In the following a cluster that contains at least one frozen sites is called frozen
cluster. In the update all clusters except for the frozen ones are flipped. 
In order to 
perform this operation only the frozen clusters have to be constructed.

Now let us discuss how $\Delta E$, eq.~(\ref{fundamental}), is computed. Let us assume 
that after equilibration we have generated a sequence of $N+1$ configurations that 
are labelled by $t$. We get 
\begin{equation}
\label{funnysum}
 \Delta E  \approx \frac{1}{N} \sum_{t=1}^N  \Delta E_{est,t}
           \approx \frac{1}{N} \sum_{t=1}^N  \frac{1}{2} (\Delta E_{est,t}+\Delta E_{est,t+1})
\end{equation}
where 
\begin{equation}
\label{naive_estimator}
 \Delta E_{est,t} = 
\sum_{<xy>} \left[s_{x,2}^{(t)} s_{y,2}^{(t)} - s_{x,1}^{(t)} s_{y,1}^{(t)} \right] 
\end{equation}
is the standard estimator of $\Delta E$. 
Combining  the right hand side of eq.~(\ref{funnysum}) and the particular properties 
of the update discussed above we arrive at the improved estimator
\begin{eqnarray}
\label{improved_estimator}
\Delta E_{imp} &=& \frac{1}{2}
 \sum_{<xy>}\left( [s_{x,2}^{(t)} s_{y,2}^{(t)} - s_{x,1}^{(t)}  s_{y,1}^{(t)}] +
                    [s_{x,2}^{(t+1)} s_{y,2}^{(t+1)} - s_{x,1}^{(t+1)} s_{y,1}^{(t+1)}] \right)
 \nonumber \\
 &=& \frac{1}{2} \sum_{<xy>} \left( [s_{x,2}^{(t)} s_{y,2}^{(t)} - s_{x,1}^{(t+1)} s_{y,1}^{(t+1)}] +
                                 [s_{x,2}^{(t+1)} s_{y,2}^{(t+1)} - s_{x,1}^{(t)} s_{y,1}^{(t)}] \right)
 \nonumber \\
&=& \frac{1}{2} \sum_{<xy> \in C_f} 
    \left( [s_{x,2}^{(t)} s_{y,2}^{(t)} - s_{x,1}^{(t+1)}  s_{y,1}^{(t+1)} ] +
              [s_{x,2}^{(t+1)} s_{y,2}^{(t+1)} - s_{x,1}^{(t)} s_{y,1}^{(t)}] \right)
\end{eqnarray}
where $<xy> \in C_f$ means that at least one of the sites $x$ or $y$ belongs to a frozen 
cluster. 
Note that for our choice of the update
\begin{equation}
 s_{x,1}^{(t+1)} s_{y,1}^{(t+1)} = s_{x,2}^{(t)} s_{y,2}^{(t)} \;\;\; \mbox{and} \;\;\;
 s_{x,2}^{(t+1)} s_{y,2}^{(t+1)} = s_{x,1}^{(t)} s_{y,1}^{(t)}
\end{equation}
for all nearest neighbour pairs $<x,y>$ where neither $x$ nor $y$ belongs to a frozen
cluster.

Whether we have achieved progress this way depends on the average 
volume taken by frozen clusters. Our numerical study discussed in detail below
shows indeed that for the whole range of $\beta$ the total size of all frozen clusters 
is small compared with the volume of the
lattice. Hence the variance of $\Delta E_{imp}$ is drastically reduced 
compared with the standard estimator. Furthermore, since the sum  runs only over the frozen 
clusters, the numerical effort to compute $\Delta E_{imp}$ is much smaller than the one 
to compute the standard estimator.

In the following we shall discuss the details of our implementation.
We simulate a large number of distances $D_{min},D_{min}+1,...,D_{max}$ jointly.
We perform the exchange cluster update and the measurement of 
$\Delta E_{imp}$ for all $D_{max}-D_{min}$ pairs $(D_1,D_2)$ with $D_2-D_1=1$.
In our C-program we store the spins in the array
\verb|int spins[I_D][L_0][L][L];| 
where \verb|I_D| equals the number of distances $D_{max}-D_{min}+1$ that are simulated. 
Routinely one would swap the spins stored in \verb|spins[i_c][i_0][i_1][i_2]|  
and \verb|spins[i_c+1][i_0][i_1][i_2]|, where the site represented by \verb|[i_0][i_1][i_2]| 
is not a member of a frozen cluster. Instead, 
in order to save CPU time, we do that for the sites that belong to frozen clusters.
This way, the systems with the distances $D_1$ and $D_2$ of the sphere from the plate
interchange their position in the array \verb|spins|. In order to keep track of where 
the systems are stored in the array \verb|spins|, we introduce the array 
\verb|int posi[I_D];| where the index \verb|i_d| equals $D-D_{min}$ and 
\verb|posi[i_d]| indicates where the system with the distance $D$ of the sphere from the 
plate is stored in \verb|spins|. Note that parallel tempering simulations  are 
usually organized in a similar way. Instead of swapping configurations, the location 
of the systems with the temperatures $T_i$ and $T_{i+1}$ is interchanged.

In order to get an ergodic algorithm we supplement the exchange cluster algorithm 
discussed above with standard updates of the individual systems. To this end we 
perform Swendsen-Wang \cite{SwWa87} updates and heat-bath updates.
In case of the Swendsen-Wang cluster-update we keep the 
clusters that contain fixed spins fixed. Those clusters that do not contain fixed spins 
are flipped with probability $1/2$.

These updates are performed in a fixed sequence: First we sweep once for each systems  
over the  whole lattice using the heat-bath algorithm.
It follows one Swendsen-Wang cluster-update for each system.
Then we perform $i_m$-times the following two steps: \\
\begin{itemize}
\item
For $D=D_{min},...,$ up to $D_{max}-1$ perform in a sequence the exchange cluster update 
and the measurement of $\Delta E_{imp}$ for the pairs of distances $D$ and $D+1$.
\item
Perform for each system one sweep with the heat-bath algorithm over a sub-lattice characterized 
by $x_0 \le l_0$, $-l_1/2 \le x_1 \le l_1/2$ and $-l_2/2 \le x_2 \le l_2/2$.
\end{itemize}
In the following we shall denote this sequence of updates as one cycle of the 
algorithm. Since the estimator $\Delta E_{imp}$ takes mainly data from the neighbourhood
of the sphere, it seems useful to perform additional updates in this region of the lattice.
Therefore we introduced the heat-bath updates of the sub-lattice. Here we made no attempt 
to optimize the shape and size of this sub-lattice nor the frequency of these updates. 

\section{Numerical Results}
\label{numerical}
As a first test of our implementation we simulated systems with the 
radius $R=2.5$  and lattices of the size  $28 \times 16^2$  and 
$38 \times 20^2$. We considered the distances $1.5 \le  D  \le 9.5$
between the sphere and the plate. We fixed $s_x=1$ for $x_0=0$ and $x_0=L_0+1$.
We simulated both $s_{sphere}=-1$ and $+1$ at the two values $\beta=0.387$
and $0.389$ of the inverse temperature. 
We simulated with the exchange cluster algorithm and without. 
For these lattice sizes we still get differences of the energies $ \Delta E$ 
that are clearly  larger than the statistical error from simulations that 
take about one day of CPU-time on a single core of the CPU in the standard
way. 
Our results with and without the  exchange cluster algorithm are compatible 
within the statistical errors. But even for the small lattices sizes used 
here, the statistical error, using an equal amount of CPU-time, is reduced by 
more than a factor of 10 by using the exchange cluster algorithm and
the improved estimator~(\ref{improved_estimator}). 
After these preliminary studies were concluded successfully, we performed an extensive 
study aiming at physically relevant results. First we summarize the parameters
of our simulations. Then we discuss the performance of the algorithm, focussing 
on the size $N_{clu}$ of the frozen clusters minus the number of fixed spins.
Finally we present our results for the thermodynamic Casimir force and compare
them with the literature. 

\subsection{Our simulations}
We studied spheres of the radii $R=3.5$, $4.5$ and $7.5$ for both
$s_{sphere}=-1$ and $1$ and $R=15.5$ for $s_{sphere}=1$ only. For these
radii, we simulated systems with $1.5 \le  D  \le D_{max}$, where 
$D_{max}=16.5$, $40.5$, $32.5$, and $10.5$ for $R=3.5$, $4.5$, $7.5$ and $15.5$, 
respectively. In the following we shall denote the number of values of $D$ by
$I_D$.  We simulated these systems for about 100 values of $\beta$.
We adapted the step-size in $\beta$. Close to $\beta_c$, the step-size is 
the smallest. For all values of $\beta$ we simulated lattices of the 
size $148 \times 80^2$, $298 \times 100^2$, $298 \times 160^2$ and 
$398 \times 200^2$, respectively.
In the case of these simulations, we fixed  $s_x=1$ for $x_0=0$ and $x_0=L_0+1$.
We fixed the parameter of the update cycle to $i_m=20 I_D$. The sizes
of the sub-lattice that is updated in excess $i_m$ times per cycle using the 
heat-bath algorithm  is 
$28  \times 16^2$, $55  \times 20^2$, $56 \times 32^2$ and $58 \times 64^2$
for $R=3.5$, $4.5$, $7.5$ and $15.5$, respectively. These parameters are
chosen ad hoc. For lack of human time we did not try to optimize them.

In order to check for the effect of the finite values of $L_0$ and $L$, we  
redid the simulations for $R=3.5$ in the neighbourhood of $\beta_c$ on lattices
of the sizes $198 \times 120^2$ and $298 \times 200^2$, where we considered both
$s_x=0$ and $s_x=1$ for $x_0=L_0+1$. For $R=4.5$ we simulated in addition 
lattices of the size $398 \times 150^2$ with $s_x=0$ for $x_0=L_0+1$ and $448 \times 150^2$
with $s_x=1$ for $x_0=L_0+1$ in the neighbourhood of $\beta_c$. For these 
additional simulations we  chose $i_m=10 I_D$. The sizes of the sub-lattice
that is updated in excess are the same as above.  In order to keep the 
figures readable we shall give in sections \ref{finieff}, \ref{kplus} and \ref{kminus}
below results for 
$s_x=1$ at $x_0=L_0+1$ only. The results obtained for $s_x=0$ at $x_0=L_0+1$
confirm our final conclusions.

Focussing on larger distances we simulated $R=3.5$ for $25.5 \le D \le 32.5$.
For  $s_{sphere}=-1$ in the range $0.3864 \le \beta \le 0.3888$ and
for $s_{sphere}=1$ in the range  $0.3864 \le \beta \le 0.3892$
we  simulated a lattice of the size $598 \times 200^2$.  
The size of the sub-lattice that is updated in excess $i_m$ times per cycle using the
heat-bath algorithm  is $43  \times 16^2$. For $\beta$ less close to $\beta_c$ we  simulated
smaller lattices. For these simulations we took $i_m=20 I_D$. 
Finally we performed simulations for $R=4.5$ at the distances $31.5 \le D \le 40.5$ using 
a lattice of the size $598 \times 200^2$  in the range $0.386 \le \beta \le 0.3886$ also 
taking $i_m=20 I_D$.

For each of these simulations we performed $10^4$ up to $2 \times 10^4$ update 
cycles. 
For $s_{sphere}=0$ and $R=3.5$, $4.5$, and $7.5$ we simulated at $\beta_c$ only.
In all our simulations we used the
Mersenne twister algorithm \cite{twister} as pseudo-random number generator.
Our code is written in C and we used the Intel compiler.
Simulating at $\beta=\beta_c$, our Swenden-Wang update takes about $1.1 \times 10^{-7}$ s
and the local heat-bath update about $3.5 \times 10^{-8}$ s per site on a single 
core of a Quad-Core AMD Opteron(tm) 2378 CPU. The exchange cluster, including the 
measurement~(\ref{improved_estimator}) takes roughly $2.6 \times 10^{-7}$ s per site 
of the frozen clusters.
In total our simulations took roughly the equivalent of 50 years of CPU time
on one core of a Quad-Core AMD Opteron(tm) 2378 CPU.

\subsection{Size of the frozen clusters}
\label{clustersize}
The CPU-time needed for one update cycle depends on the size of the frozen clusters.
Also the variance of the estimator~(\ref{improved_estimator}) compared with the 
standard one can be small only if the size of the frozen clusters is small compared 
with the volume of the lattice. Therefore we shall discuss the  size of the frozen 
clusters and its dependence on the parameters $\beta$, $R$, $D$ and the lattice size
in some detail below.

To give a first impression, we plot
\begin{figure}
\begin{center}
\includegraphics[width=13.5cm]{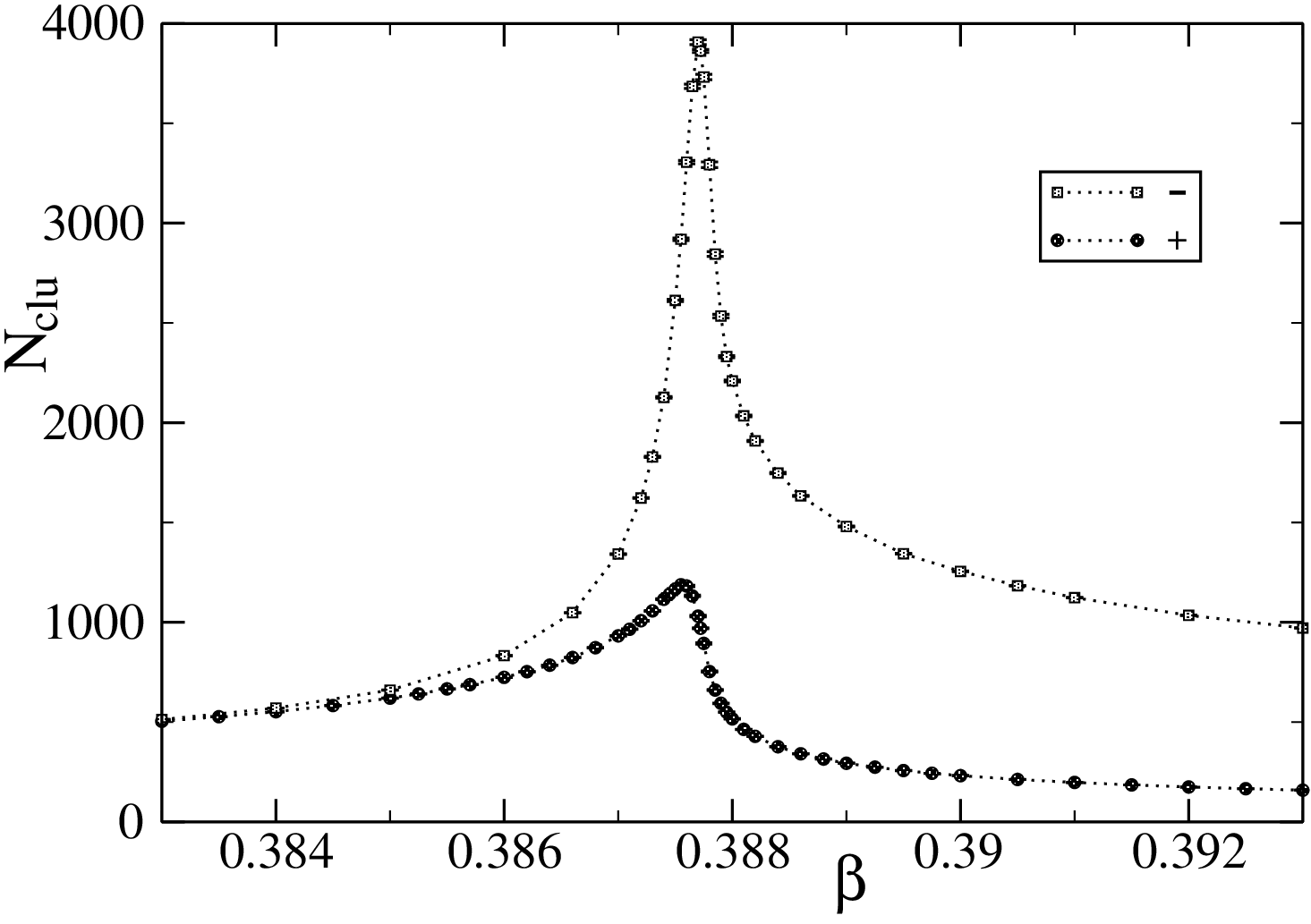}
\caption{\label{ball31}
We plot the average size $N_{clu}$ of the frozen clusters minus the number of frozen 
sites for the radius $R=7.5$ and the distance $D=32$ as a function of the inverse temperature $\beta$.
The upper and the lower curve represent the results for $s_{sphere}=-1$ and $1$, respectively.
}
\end{center}
\end{figure}
in Fig.  \ref{ball31} the average size $N_{clu}$ of the frozen clusters 
minus the number of frozen sites for the radius $R=7.5$ and the distance $D=32$,
i.e. for the pair of systems with $D=31.5$ and $32.5$,  
for $s_{sphere}=-1$ and $1$. For decreasing $\beta$ in the high 
temperature phase, the average size of the frozen clusters for 
$s_{sphere}=-1$ and $1$ seems to become identical.
As the critical point is approached, the size for $s_{sphere}=-1$
becomes clearly larger than for $s_{sphere}=1$. This remains true
in the low temperature phase. In both cases, the maximum $N_{clu,max}$
is located close to $\beta_c$. We find $N_{clu,max} \approx 3900$ and
$N_{clu,max} \approx 1190$ for $s_{sphere}=-1$ and $1$, respectively.
Note that this is only a small fraction of the total volume of the lattice,
$298 \times 160^2=7628800$.  

Let us now discuss the behaviour of $N_{clu}$ in more detail.
Since $p_d = 1$ for $\beta = 0$, the frozen clusters consist of the frozen 
sites only, i.e. $N_{clu}=0$. At low temperatures, the configurations become ordered. 
Therefore also in this limit the frozen clusters consist of the frozen sites
only.

From our numerical data we see that for fixed $\beta$ and lattice size, $N_{clu}$ is monotonically
increasing with increasing distance $D$. For $L_0, L, D \gg \xi$,  $N_{clu}$ seems to 
converge to a finite value $N_{clu,\infty}$. In the high temperature phase, $\beta < \beta_c$, 
this value is the same for $s_{sphere}=-1$ and $1$. We fitted our data in the high temperature 
phase with the Ansatz
$N_{clu,\infty} = c (\beta_c-\beta)^{-y}$. For $R=3.5$ we get quite reasonable fits
with the result $y \approx 0.6$.  Fits for larger radii, where we have less data  apparently
give smaller values of $y$. However, since $N_{clu,\infty}$ should increase with increasing radius $R$,
this observation seems to be an artifact of our finite data set. 
A theoretical understanding of the behaviour of $N_{clu,\infty}$, relating $y$ with known critical
exponents, analogous to ref. \cite{HeBl98}, would be desirable.

Next let us study the dependence of $N_{clu}$ on the linear lattice sizes $L_0$ and $L$ in the 
\begin{figure}
\begin{center}
\includegraphics[width=13.5cm]{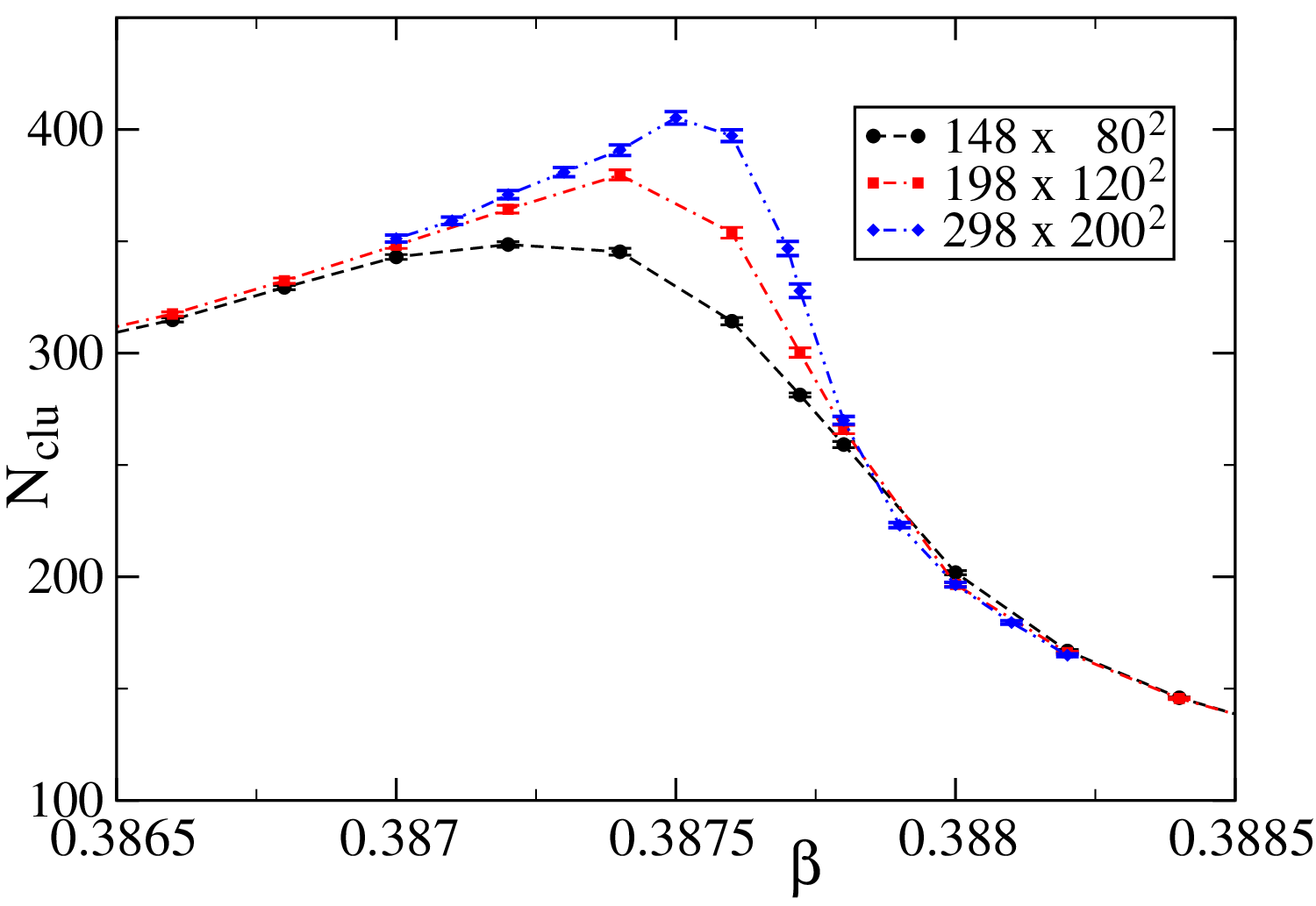}
\caption{\label{finite4}
We plot  $N_{clu}$ as a function of $\beta$ for $R=3.5$, $D=16$ and $s_{sphere}=1$  focussing on 
the neighbourhood
of the critical point. We give results for the lattices sizes $148 \times 80^2$, $198 \times 120^2$
and $298 \times 200^2$.
}
\end{center}
\end{figure}
neighbourhood of the critical point. In Fig. \ref{finite4} we plot $N_{clu}$ as a function 
of $\beta$ for $R=3.5$, $D=16$  and $s_{sphere} = 1$ for the lattice sizes $148 \times 80^2$, 
$198 \times 120^2$ and $298 \times 200^2$.
For  $\beta \lessapprox  0.387$ and $\beta \gtrapprox  0.3882$ the estimates of $N_{clu}$ for
the different lattice sizes are consistent within the statistical error.  Instead for 
$0.387  \lessapprox \beta \lessapprox  0.3882$ a dependence of $N_{clu}$ on the lattice size 
can be seen. For most values of $\beta$ in this range $N_{clu}$ is increasing with increasing lattice 
size. However for $\beta \approx 0.388$ the value of $N_{clu}$ for $148 \times 80^2$ seems to be 
a bit larger than those for  the lattices sizes $198 \times 120^2$ and $298 \times 200^2$. 
For $s_{sphere} = -1$ we get a qualitatively similar behaviour. The same holds for $R=4.5$ for 
both $s_{sphere} = -1$ and $1$.

\begin{figure}
\begin{center}
\includegraphics[width=13.5cm]{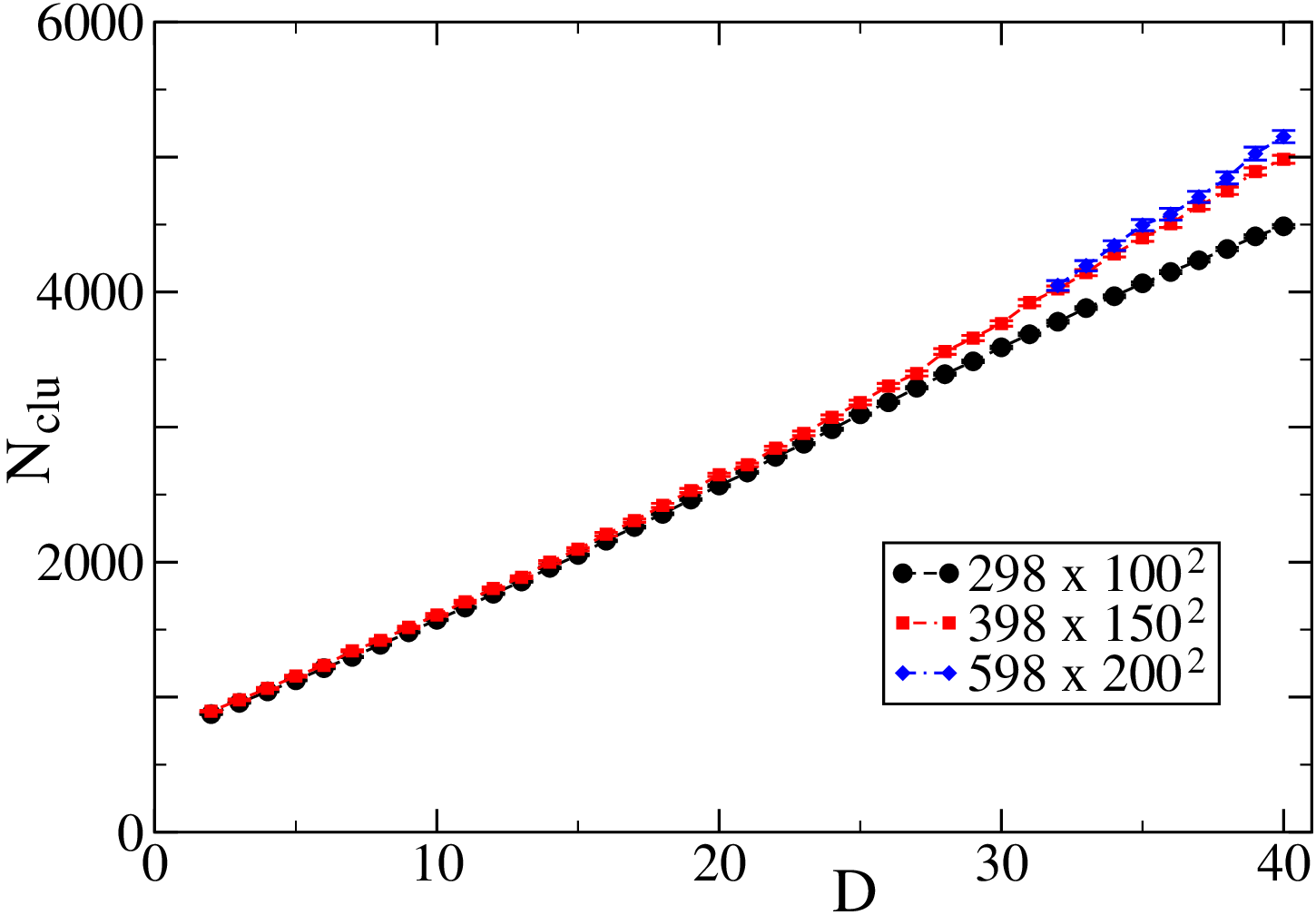}
\caption{\label{betac5m}
We plot  $N_{clu}$ as a function of $D$ for $R=4.5$,  $s_{sphere}=-1$ and
$\beta=\beta_c$.
We give results for the lattice sizes $298 \times 100^2$, $448 \times 150^2$
and $598 \times 200^2$.
}
\end{center}
\end{figure}

Finally let us discuss the behaviour of $N_{clu}$ at the critical point in 
more detail. In Fig. \ref{betac5m} we plot $N_{clu}$ as a function of $D$ 
for $R=4.5$,  $s_{sphere}=-1$ and $\beta=\beta_c$. The value of $N_{clu}$ 
seems to increase somewhat faster than linearly with the distance $D$. 
Differences between the 
results for the lattice sizes $298 \times 100^2$, $448 \times 150^2$
and $598 \times 200^2$ are clearly visible.  Similar observations 
can be made for $s_{sphere}=1$ and also for other radii.

In table \ref{Csize} we give $N_{clu}$ at $\beta_c$ for the radii $R=3.5$, $4.5$ and $7.5$ for 
\begin{table}
\caption{\sl \label{Csize}  $N_{clu}$ for $s_{sphere}=-1$, $0$ and
$1$ at the critical point for $D \approx 4.3 R$.  The results
are taken from the lattice sizes $148 \times 80^2$, $298 \times 100^2$,
$298 \times 160^2$  for $R=3.5$, $4.5$ and $7.5$, respectively.
}
\begin{center}
\begin{tabular}{lccc}
\hline
  $R$ $\diagdown$ $s_{sphere}$  &  $-1$ & $0$ & $1$  \\
\hline
  3.5  & 1096(3)& 1495(5)  & 269(1) \\
  4.5  & 2464(6)& \phantom{0}4201(19) & 565(2) \\
  7.5  & 3862(8)& 11203(67)  & 970(2) \\
\hline
\end{tabular}
\end{center}
\end{table}
the distance $D \approx 4.3 R$, i.e. $D=15$, $19$, and  $32$, respectively.  
As one might expect, $N_{clu}$ is increasing with increasing radius $R$. 
The ratio of
$N_{clu}$ for $s_{sphere}=-1$ and $s_{sphere}=1$ is roughly the same for all three radii. 
In contrast, we find that $N_{clu}$ is increasing more rapidly with increasing 
$R$ for $s_{sphere}=0$ than for $s_{sphere}=\pm 1$.  For $s_{sphere}=\pm 1$ the results
for $R=4.5$ and $7.5$ suggest that $N_{clu}$ is increasing less than
linearly with the radius $R$, while the opposite seems to be true for $s_{sphere}=0$. 

In section \ref{algorithm}  we argued that the estimator~(\ref{improved_estimator}) reduces
the variance if $N_{clu} \ll L_0 \times L^2$. For $s_{sphere}=\pm 1$ we have verified that 
this is indeed the case for the full range of $\beta$ and all radii and lattice sizes
studied here. Our preliminary results for $s_{sphere}=0$  obtained at $\beta_c$ indicate that
in this case the exchange cluster algorithm is less efficient than for $s_{sphere}=\pm 1$. 
Still it 
seems plausible that also for $s_{sphere}=0$ meaningful results for the thermodynamic
Casimir force can be obtained.

\subsection{The thermodynamic Casimir force: Effects of the finite lattice}
\label{finieff}
\begin{figure}
\begin{center}
\includegraphics[width=12.4cm]{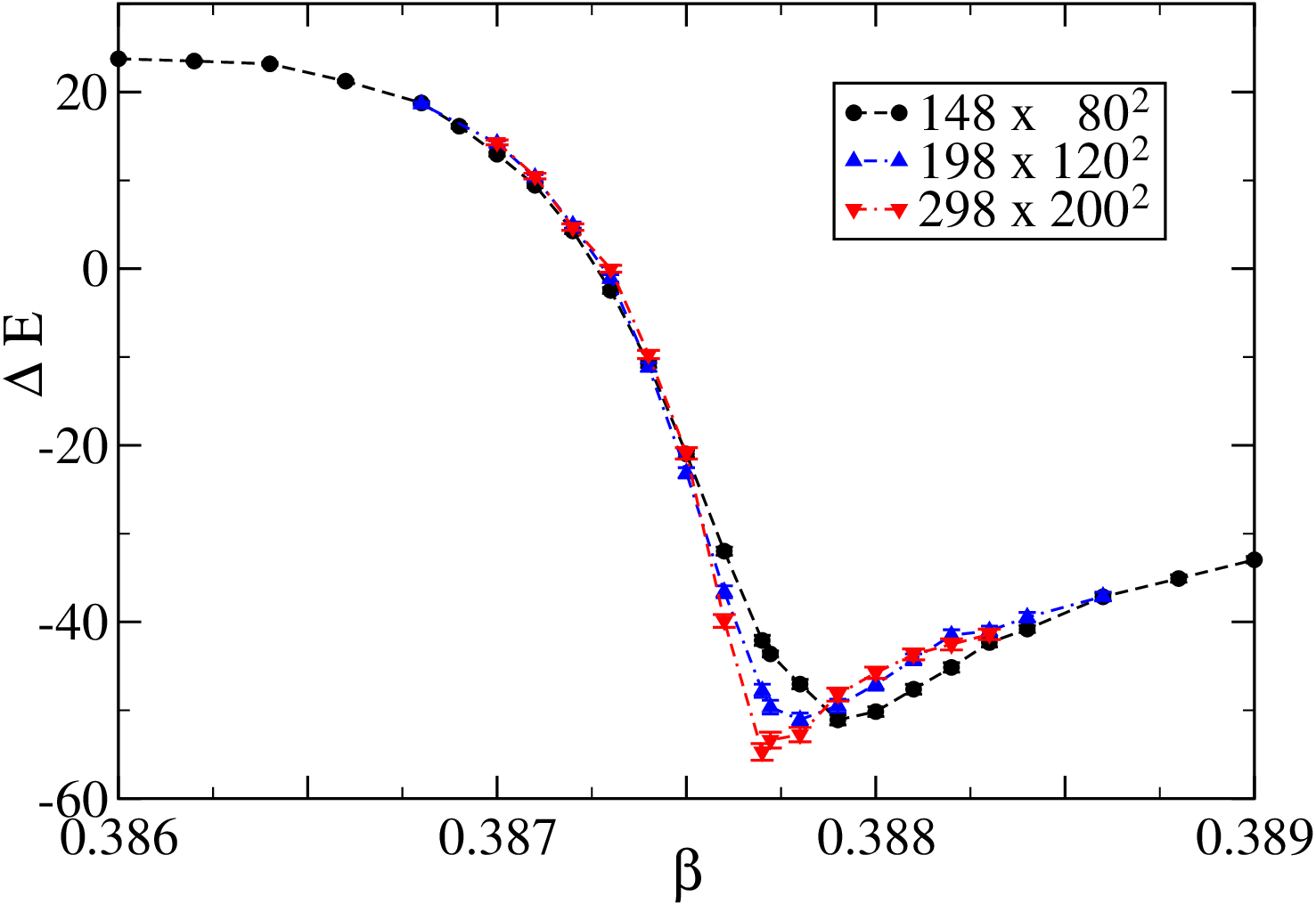}
\vskip1.7cm
\includegraphics[width=12.4cm]{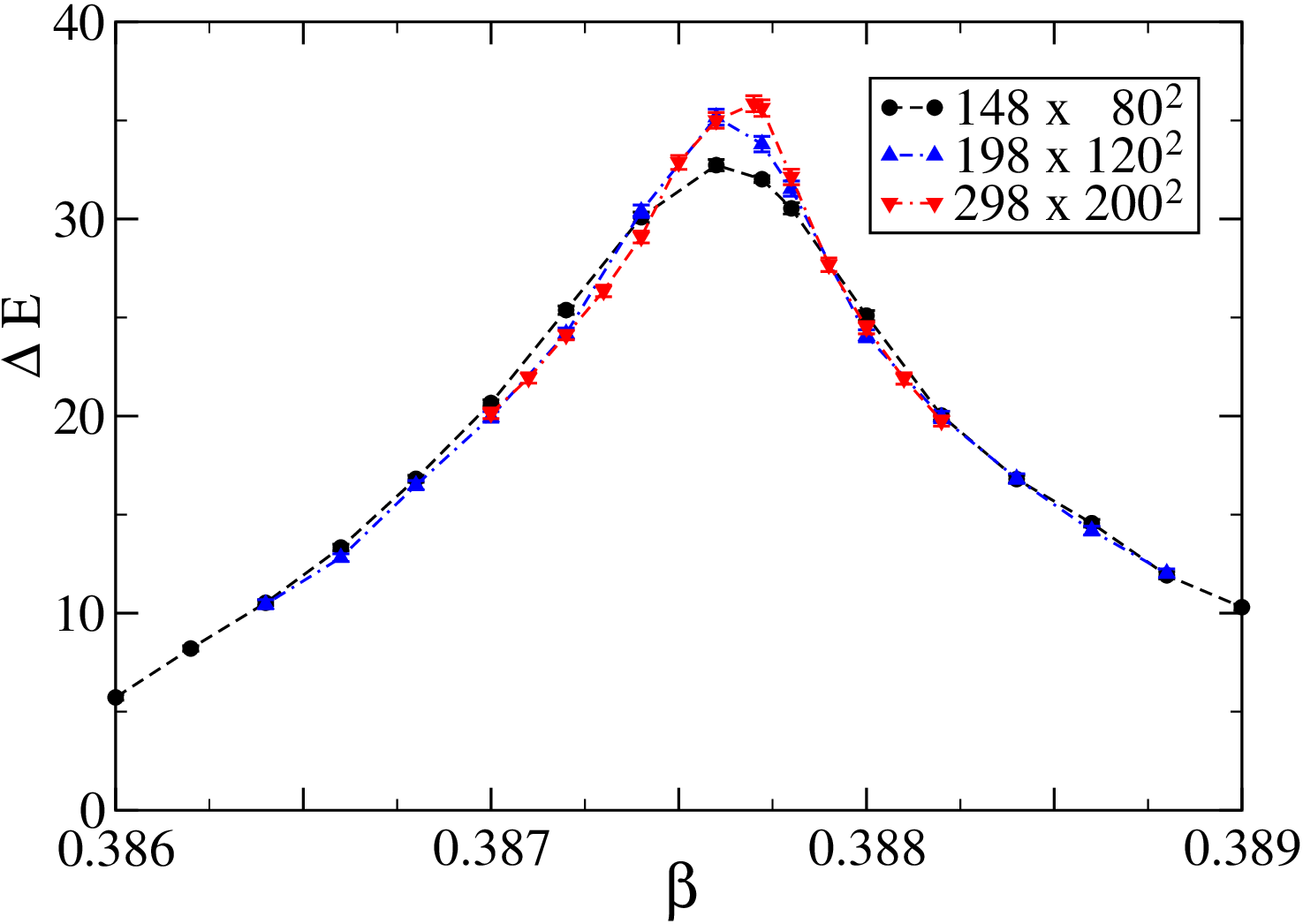}
\caption{\label{finite4B}
We plot $\Delta E$ as a function of $\beta$ for $R=3.5$ and $D=16$. The 
lattice sizes are $148 \times 80^2$, $198 \times 120^2$, and $298 \times 200^2$.
In the upper and the lower part of the figure we give results for $s_{sphere}=-1$
and $s_{sphere}=1$, respectively.
}
\end{center}
\end{figure}
For $\beta \ne \beta_c$  corrections to the limit 
$L_0,L \rightarrow \infty$ decay exponentially  with increasing lattice size.
Therefore corrections should be negligible for $L_0,L \gg \xi_{bulk}$. 
Instead, at the critical point, we expect that corrections decay power-like.
For $R=3.5$ we checked explicitly how $\Delta E$ depends 
on the lattice size. To this end we simulated in the neighbourhood of $\beta_c$
in addition to the lattice of the size $148 \times 80^2$  ones
of the sizes $198 \times 120^2$ and $298 \times 200^2$.
In Fig. \ref{finite4B} we plot for $D=16$ our results 
for the difference of energies $\Delta E$ as a function of $\beta$.
We give our results for $s_{sphere} = -1$ and $1$ in the 
upper and the low part of Fig.  \ref{finite4B}, respectively. We see differences 
between the results obtained for the $148 \times 80^2$ and the larger lattices
that are statistically significant  for $0.3868 \lessapprox  \beta \lessapprox 0.3886$
and $0.3868 \lessapprox  \beta \lessapprox 0.3882$ for $s_{sphere} = -1$ and $1$, 
respectively.  Note that $\xi_{bulk}(0.3868) \approx 18.66$, using eq.~(\ref{xi0}).

\begin{figure}
\begin{center}
\includegraphics[width=12.4cm]{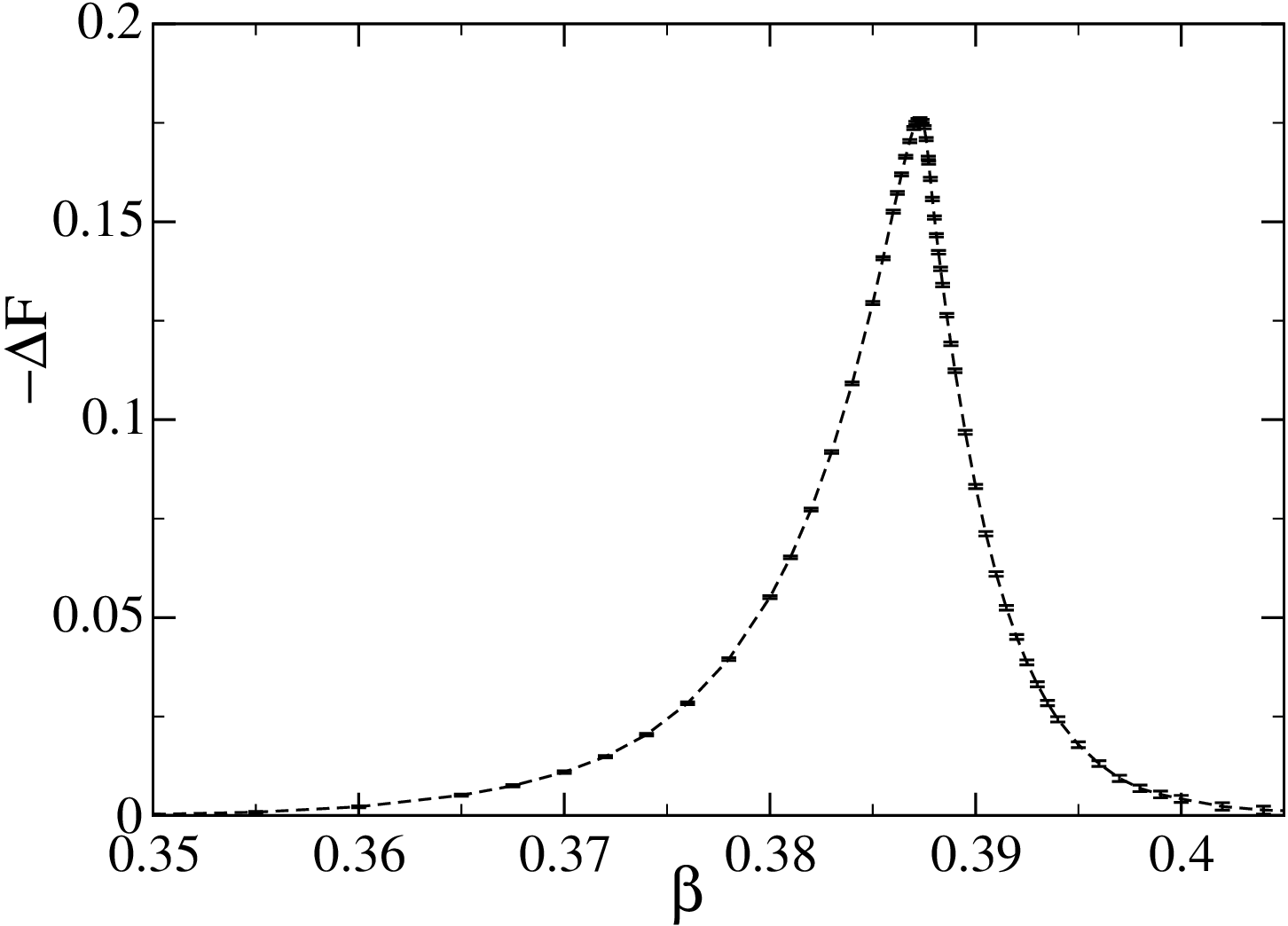}
\vskip2.3cm
\includegraphics[width=12.4cm]{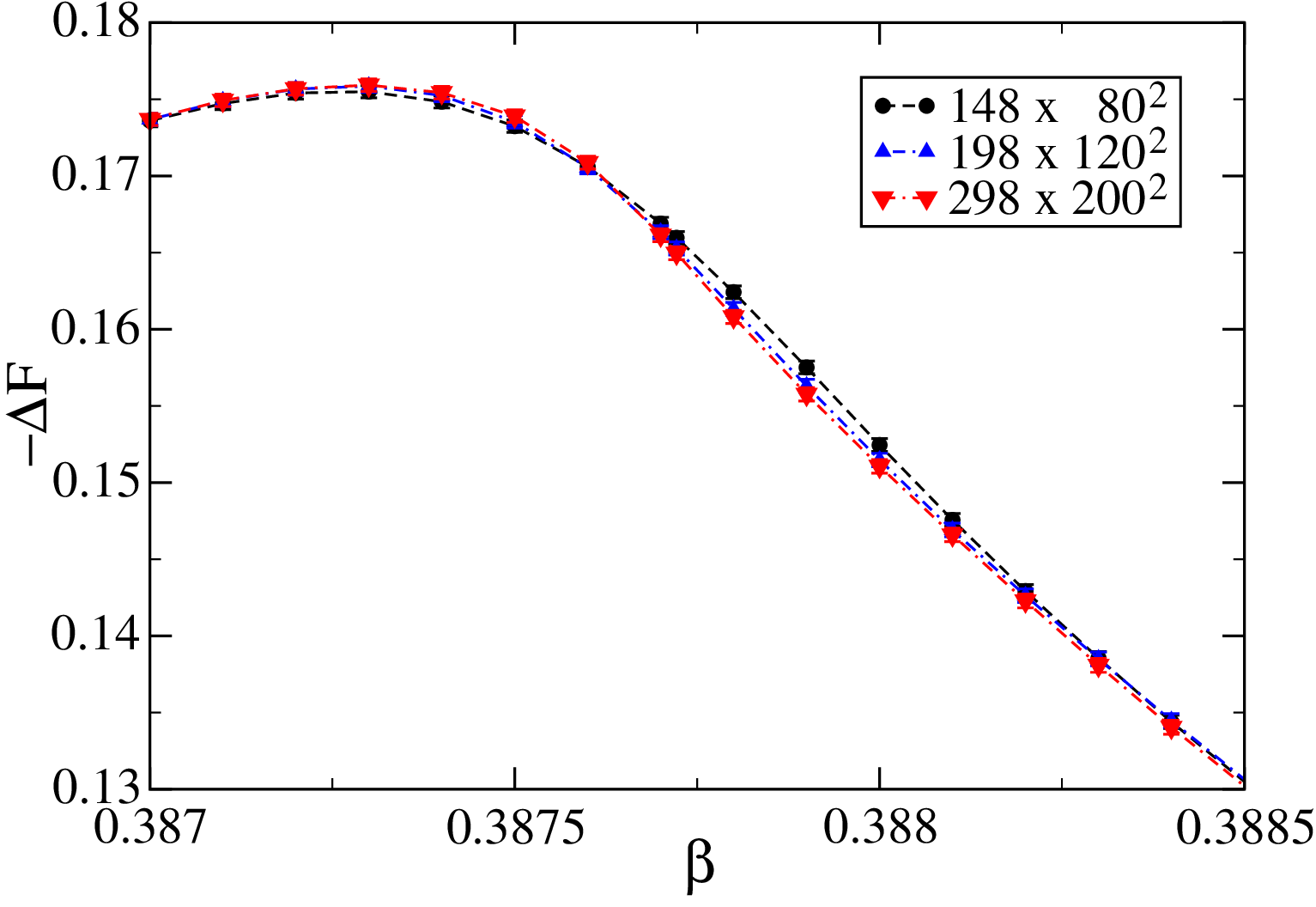}
\caption{\label{finite4mint}
We plot $-\Delta F$ as a function of $\beta$ for $R=3.5$,  
$D=16$, and $s_{sphere}=-1$. In the upper part of the figure we give the result obtained by
using a $148 \times 80^2$ lattice for the full range of $\beta$ that 
we consider. In the lower part of the figure we focus on the 
neighbourhood of $\beta_c$,  plotting the results for the 
lattice sizes  $148 \times 80^2$, $198 \times 120^2$, and $298 \times 200^2$.
}
\end{center}
\end{figure}

In Fig. \ref{finite4mint} we give the numerical results of $-\Delta F$
for $R=3.5$, $D=16$ and $s_{sphere}=-1$. In the upper part of the
figure we give the result of the numerical integration of $\Delta E$ obtained
by using the data for the $148 \times 80^2$ lattice. In the lower part of the figure 
we focus on the neighbourhood of $\beta_c$, where effects due to finite
$L_0$ and $L$ become important. As check, we computed $-\Delta F$ 
in an alternative way: We replaced the estimates of $\Delta E$ from 
the $148 \times 80^2$ lattice, where available, by those from the 
$198 \times 120^2$ or the $298 \times 200^2$ lattice. We find that  
the behaviour of $-\Delta F$ is changed only little. The difference between
these results is of a similar size as the statistical error. Note that
the statistical error of $-\Delta F$ is the accumulated error of $\Delta E$
over the range of the integration.  

We conclude that in a neighbourhood of the critical point 
the effect of the finite lattice size on $\Delta E$ is clearly visible.
However, since the range of $\beta$ where this is the case, is quite 
small, $-\Delta F$ is affected less. The amplitude 
of this effect can be estimated by comparing the results obtained
for different lattice sizes.

\subsection{The scaling function $K_+(x,\Delta)$ for $s_{sphere}=1$}
\label{kplus}
 In this section we discuss our results for the scaling function $K_+(x,\Delta)$ for 
$s_{sphere}=1$. In a first step we check how well our results obtained 
for different radii collapse on a single curve. To this end we make use of the 
effective radius determined in Appendix \ref{App1}, see table 
\ref{refftab2}, and the effective distance~(\ref{Deff}).  In particular
$\Delta = D_{eff}/R_{eff}$ in the following analysis of our data.  In the 
plots given below we show the statistical error of the data only. The 
uncertainty of $D_{eff}$ and $R_{eff}$ still comes on top of this.
Essentially it is a two percent uncertainty of the scales.
\begin{figure}
\begin{center}
\includegraphics[width=12.3cm]{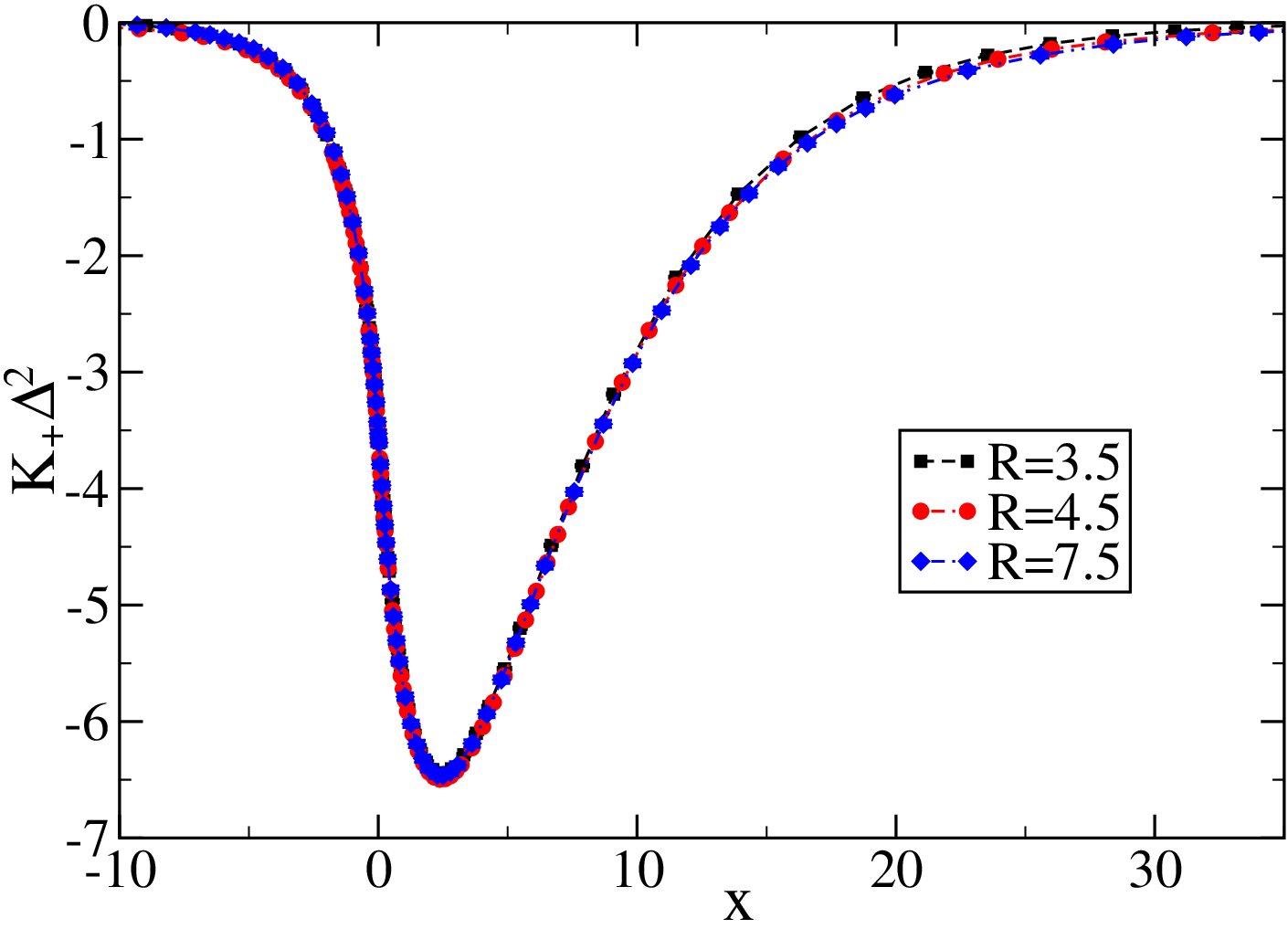}
\vskip2.3cm
\includegraphics[width=12.3cm]{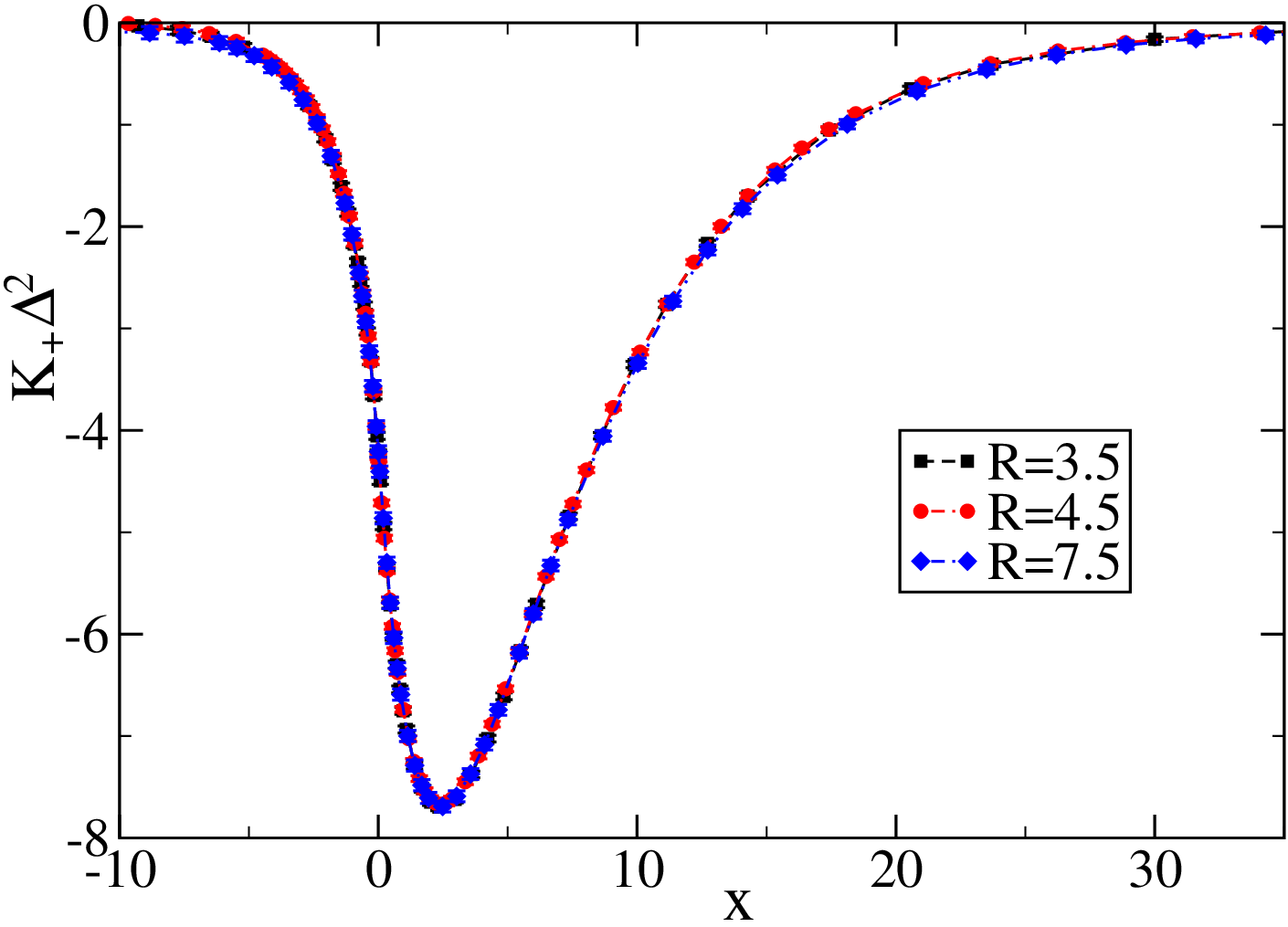}
\caption{\label{scaling2}
We plot $-R_{eff} \Delta F \Delta^2$ as a function of $x=t [D_{eff}/\xi_0]^{1/\nu}$ for $s_{sphere}=1$
and  the radii $R=3.5$, $4.5$  and  $R=7.5$. 
 In the upper part we give results for $\Delta \approx 2.7$, 
which means $D=6$, $9$ and $18$ for $R=3.5$, $4.5$ and $7.5$, respectively.
In the lower part we give results for $\Delta \approx 4.8$,
which means $D=12$, $17$ and $32$ for $R=3.5$, $4.5$ and $7.5$, respectively. Note that
$\Delta=D_{eff}/R_{eff}$ is used.
}
\end{center}
\end{figure}

In figure \ref{scaling2}  we plot $-R_{eff} \Delta F \Delta^2$ as a function of 
$x=t [D_{eff}/\xi_0]^{1/\nu}$ for $\Delta \approx 2.7$ and $4.8$. We multiplied 
by $\Delta^2$ to reduce the dependence on $\Delta$. Note that for small 
values of $\Delta$, $K(0,\Delta) \propto \Delta^{-2}$ and large ones
$K(0,\Delta) \propto \Delta^{-\beta/\nu-1}$ \cite{HaScEiDi98,BuEi97}, where 
$\beta/\nu=(1+\eta)/2$ and $\eta=0.03627(10)$ \cite{mycritical}.  We took
the data obtained by using $148 \times 80^2$, $298 \times 100^2$ and 
$298 \times 160^2$ lattices for the radii $R=3.5$, $4.5$ and $7.5$, respectively.
For this choice $L/R$ is roughly the same for the three radii considered.

In particular for $\Delta \approx 4.8$ we see an excellent collapse of the 
data obtained for the three different radii. Note that things look far worse
when $R$ and $D$ are used instead of $R_{eff}$ and $D_{eff}$.  For 
$\Delta \approx 2.7$ the collapse of the data is slightly worse than for $\Delta 
\approx 4.8$. In particular for  $x \gtrapprox 10$ we see some deviation of 
the result for $R=3.5$ from that for the two other radii. This is not too
surprising, since $\Delta \approx 2.7$ means $D=6$ for $R=3.5$ and also the values 
of $\beta$ that correspond to $x \gtrapprox 10$ are far from $\beta_c$.

We conclude that the data scale well and our numerical results should 
describe the scaling limit. At the level of our accuracy, one should 
interpret  data obtained for $D \lessapprox 6$ with caution.

\begin{figure}
\begin{center}
\includegraphics[width=15.3cm]{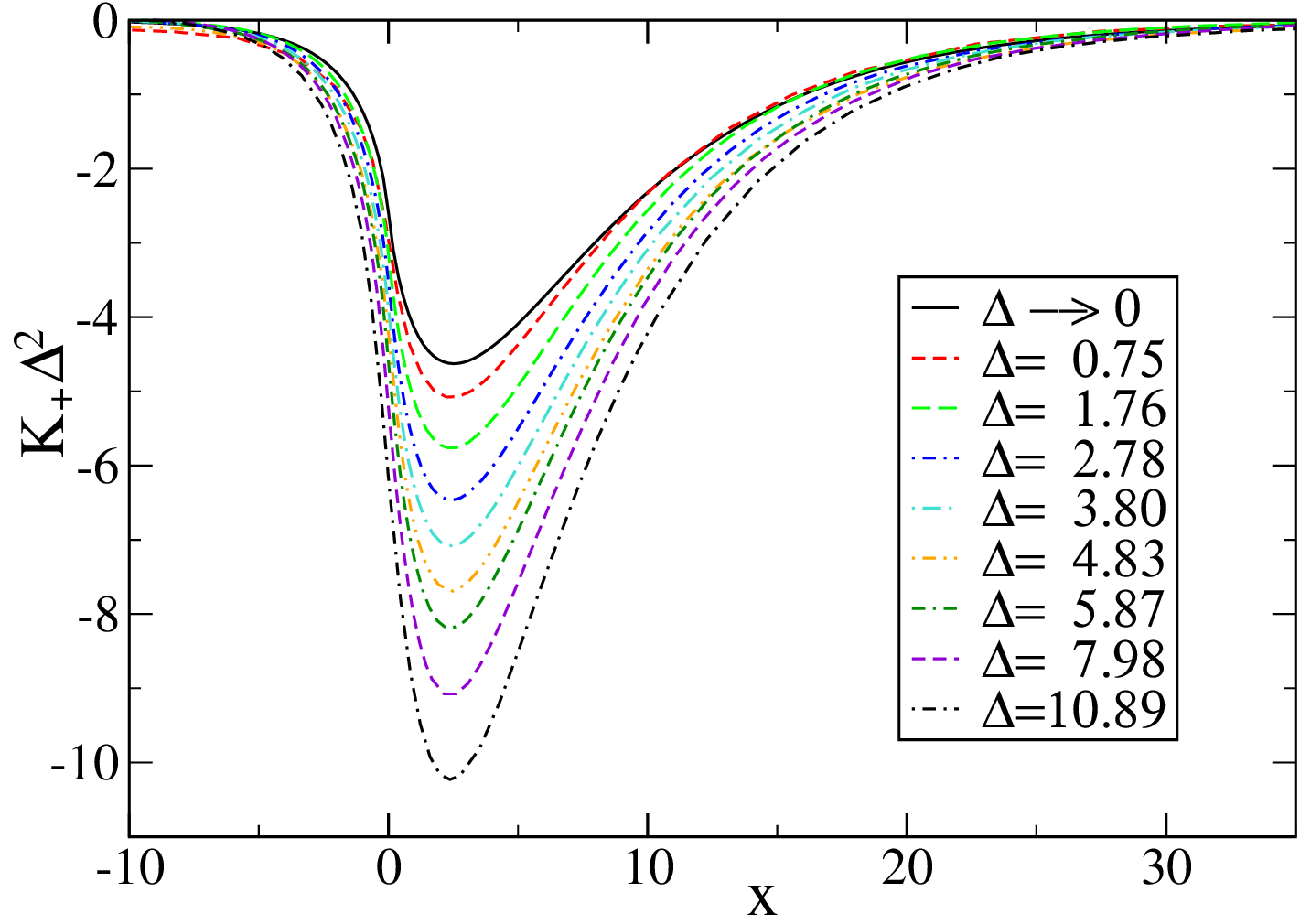}
\caption{\label{manyp}
We plot $K_+(x,\Delta) \Delta^2$ as a function of $x$ for various values of $\Delta$.
Note that the index of $K$ indicates $s_{sphere}=1$.
We compare the Derjaguin approximation in the limit $\Delta \rightarrow 0$
with our Monte Carlo data for various finite values of $\Delta$.
In particular we have used our data for $R=15.5$ and $D=10$ giving 
$\Delta \approx 0.75$,  $R=7.5$ and $D=11$, $18$, $25$, and $32$ giving $\Delta \approx 1.76$, 
$2.78$, $3.80$, and $4.83$, respectively, and $R=4.5$ and $D=21$, $29$ and  $40$ giving 
$\Delta \approx 5.87$, $7.98$,  and  $10.89$, respectively.
}
\end{center}
\end{figure}

In Fig. \ref{manyp} we plot $K_+(x,\Delta) \Delta^2$ as a function of 
$x=t [D/\xi_0]^{1/\nu}$ for various
values of $\Delta$ in the range $0.75 \leq \Delta \leq 10.89$. For each 
value of $\Delta$, we took the largest radius, where data are available. 
For comparison 
we plotted the  Derjaguin approximation in the limit $\Delta \rightarrow 0$.
For a brief discussion of the Derjaguin approximation see e.g. ref. \cite{HaScEiDi98}. 
As input we used the results for $\theta_{++}(x)$ and $\theta_{+-}(x)$ obtained in 
ref. \cite{mycorrection}.

We find that the shape of $K_+(x,\Delta) \Delta^2$ changes only little with 
varying $\Delta$. In particular the location of the minimum changes very little 
with varying $\Delta$.  The absolute value of $K_+(x,\Delta) \Delta^2$ is increasing 
with increasing $\Delta$, which is consistent with the fact that for large
$\Delta$, $K_+(0,\Delta) \propto \Delta^{-\beta/\nu-1}$ \cite{BuEi97}. These observations are
fully consistent with those of ref. \cite{troendle}. In Fig. 2 a of \cite{troendle} 
$K_{(\pm,-)}(\theta,\Delta)/|K_{(\pm,-)}(0,\Delta)|$ is plotted as a function of 
$\theta=\mbox{sign}(t) D/\xi$. Note that the index $(-,-)$ corresponds to our $+$ and 
$(+,-)$ to our $-$.  For $(-,-)$, the mean-field result for $\Delta=1/3$ is fully consistent with the
Derjaguin approximation for four dimensions, while for $\Delta=1$ a small deviation
can be seen.

\begin{figure}
\begin{center}
\includegraphics[width=14.5cm]{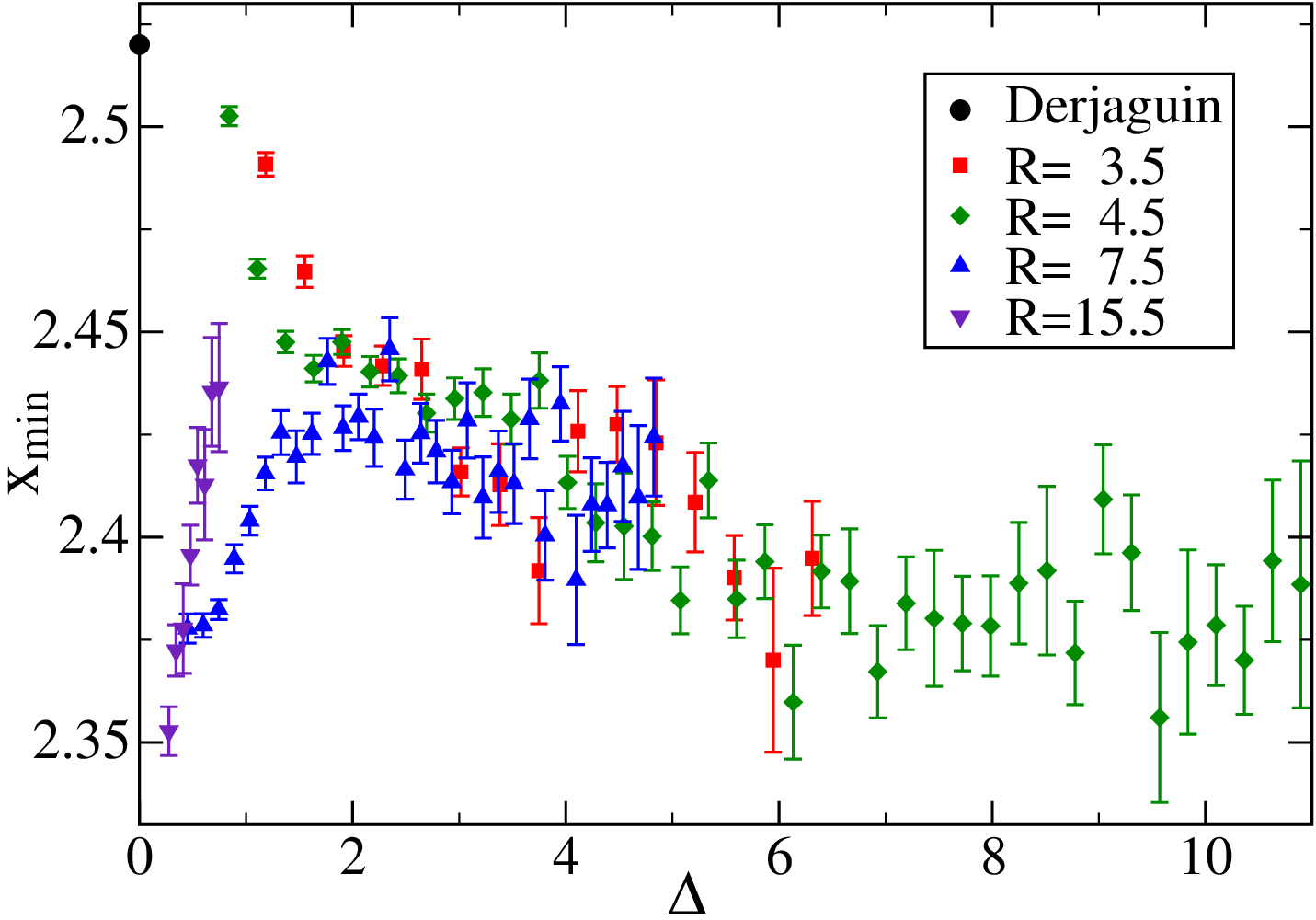}
\caption{\label{xmin}
We plot our estimates of $x_{min}$ as a function of $\Delta$. Monte Carlo results obtained by using 
$R=3.5$, $4.5$, $7.5$ and $15.5$ are given. These are compared with the Derjaguin approximation 
in the limit $\Delta \rightarrow 0$.  Note that $s_{sphere}=1$.
}
\end{center}
\end{figure}

In the following we study in more detail how the minimum of $K_+(x,\Delta)$  as 
function of $x$ depends on $\Delta$.  Fortunately, $x_{min}$ is  large enough, 
such that finite $L_0$ and $L$ effects are well under control.
We located the minimum of $-\Delta F$ by finding the zero of $\Delta E$. To this end, 
we interpolated $\Delta E$ linearly in $\beta$.  Using these results, we computed 
$x_{min} = (\beta_c-\beta_{min}) [D_{eff}/\xi_0]^{1/\nu}$. 
In Fig. \ref{xmin} we plot $x_{min}$ as a function of $\Delta$ using our data obtained 
for the radii $R=3.5$, $4.5$, $7.5$ and $15.5$.  We see that for $\Delta \gtrapprox 2$ the results 
obtained for different radii agree quite nicely. At $\Delta=2$ the estimate $x_{min} \approx 2.43$ seems
plausible. With further increasing $\Delta$ the value of $x_{min}$ is slightly decreasing. For 
$\Delta \approx 12$ we get  $x_{min} \approx 2.3$.   For $\Delta \lessapprox 2$ clear deviations 
between the results obtained by using different radii $R$ can be observed. Therefore we abstain 
to give a conclusion for the scaling limit in this range. Note that the Derjaguin approximation gives
$x_{min} \approx 2.52$ for the limit $\Delta  \rightarrow 0$. The Derjaguin approximation predicts
that $x_{min}$ is increasing with increasing distance $\Delta$.  If that prediction is correct, $x_{min}$
should show a maximum in the interval $0 < \Delta \lessapprox 1$. 

Our results can be compared with the upper part of FIG. 2 of ref. \cite{HaScEiDi98}.  
Note that the authors 
of ref. \cite{HaScEiDi98} consider $D/\xi \simeq x^{\nu}$ as variable.  They obtain 
$x_{min} = 1.6^{1/0.63002} = 2.1...$ for $\Delta=0$ using the Derjaguin approximation.  This discrepancy 
should be due to the different numerical estimates of the scaling function $\theta_{++}$
of the plate-plate geometry. E.g. following FIG 3 of ref. \cite{HaScEiDi98}
$\theta_{++}(0)=-0.326$, while our present estimate is $\theta_{++}(0) = - 0.410(8)$ \cite{mycorrection}. 
Then, according to ref. \cite{HaScEiDi98} $x_{min}$ increases and reaches a maximum at $\Delta \approx 3$ 
assuming the value $x_{min} \approx 2.5^{1/0.63002} \approx 4.3$.
Then, with increasing $\Delta$ it is decreasing
and reaches $x_{min} \approx 1.5^{1/0.63002} \approx 1.9$ in the limit $\Delta \rightarrow \infty$. 
In particular for the value assumed at $\Delta \approx 3$ we see a clear discrepancy with 
our results. It is beyond the scope of the present work to discuss the reliability of the first order
$\epsilon$-expansion that has been used in ref. \cite{HaScEiDi98} to get the small-sphere expansion.

\begin{figure}
\begin{center}
\includegraphics[width=14.5cm]{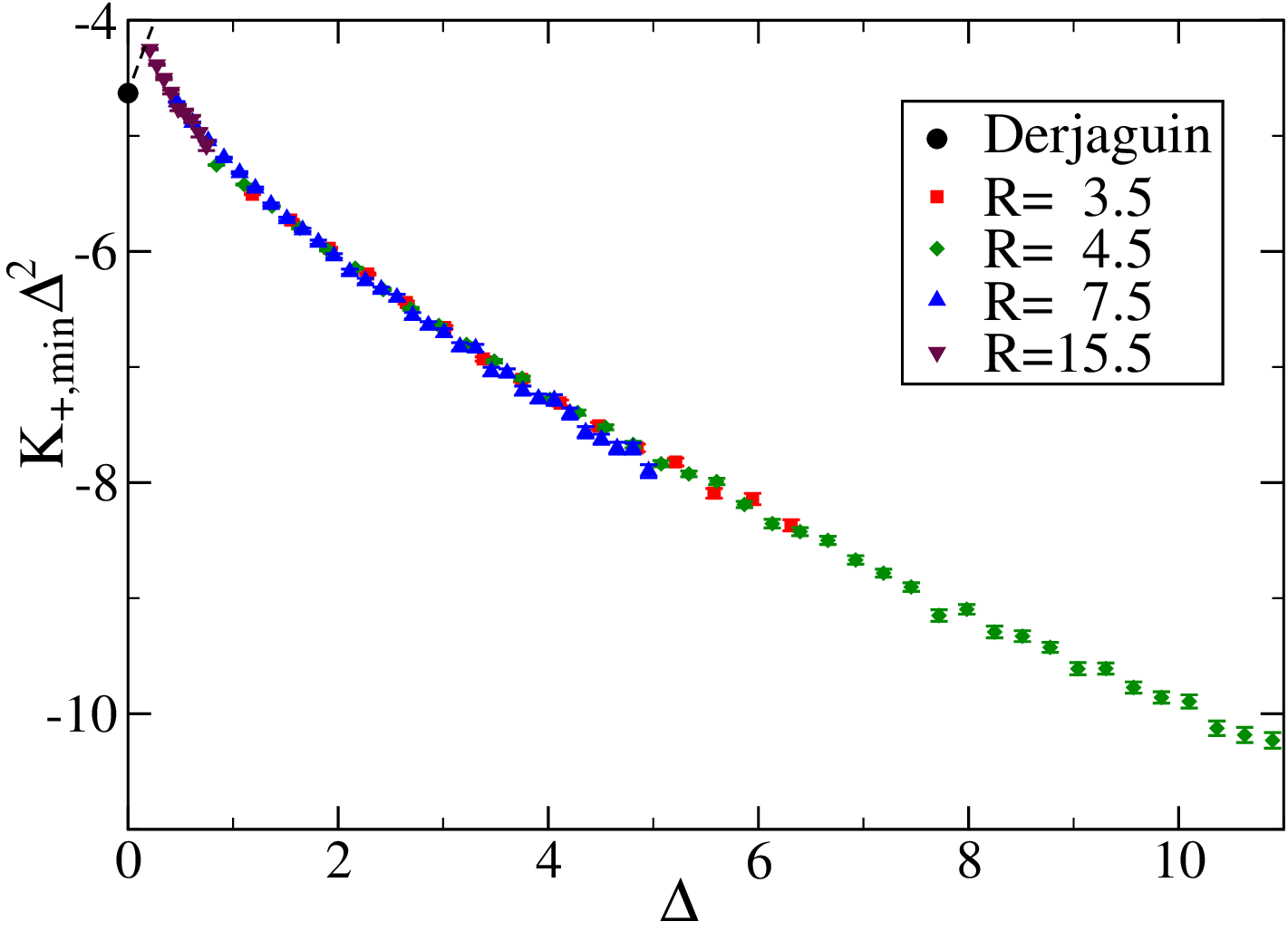}
\caption{\label{min2}
We plot  our estimates of  $K_+(x_{min},\Delta)  \Delta^2$ as a function of $\Delta$. 
Monte Carlo results obtained by using $R=3.5$, $4.5$, $7.5$ and $15.5$ are given.
The Derjaguin approximation in the limit $\Delta \rightarrow 0$ is given by a black circle. 
The Derjaguin approximation for finite values of $\Delta$ is represented by a dashed black line.
Note that $s_{sphere}=1$.
}
\end{center}
\end{figure}

In Fig. \ref{min2} we plot  $K_+(x_{min},\Delta) \Delta^2$ as a function of $\Delta$.  We find a nice
collapse of the data obtained from different radii of the sphere. In contrast to $x_{min}$, this is also
the case for $\Delta \lessapprox 2$.  Taking into account our Monte Carlo data only, 
$K_+(x_{min},\Delta) \Delta^2$ seems to be a monotonically decreasing function of $\Delta$.  
However we note
that for $\Delta \lessapprox 0.4$ our numerical estimates of $K_+(x_{min},\Delta) \Delta^2$ 
are overshooting  
$\lim_{\Delta \rightarrow 0} K_+(x_{min},\Delta) \Delta^2 \approx -4.63$ obtained from the Derjaguin 
approximation. Also note that the Derjaguin approximation predicts that $K_+(x_{min},\Delta) \Delta^2$
increases with increasing $\Delta$, which is indicated by a dashed black line in Fig.  \ref{min2}.  
Therefore $K_+(x_{min},\Delta) \Delta^2$ should  show a maximum in the interval 
$0 < \Delta \lessapprox 0.4$.
In the lower part of Fig. 2. of ref. \cite{HaScEiDi98} $K_+(x_{min},\Delta) \Delta^{1+\beta/\nu}$ is shown 
as a function of $\Delta$. For $\Delta \approx 10$ it shows a maximum (minimum with the sign of
ref. \cite{HaScEiDi98} ) taking a value slightly larger than $-3$. Taking our Monte Carlo data, 
$K_+(x_{min},\Delta) \Delta^{1+\beta/\nu}$ is still slightly increasing at $\Delta \approx 10$ and takes
the value $\approx -3.25$.

\begin{figure}
\begin{center}
\includegraphics[width=14.3cm]{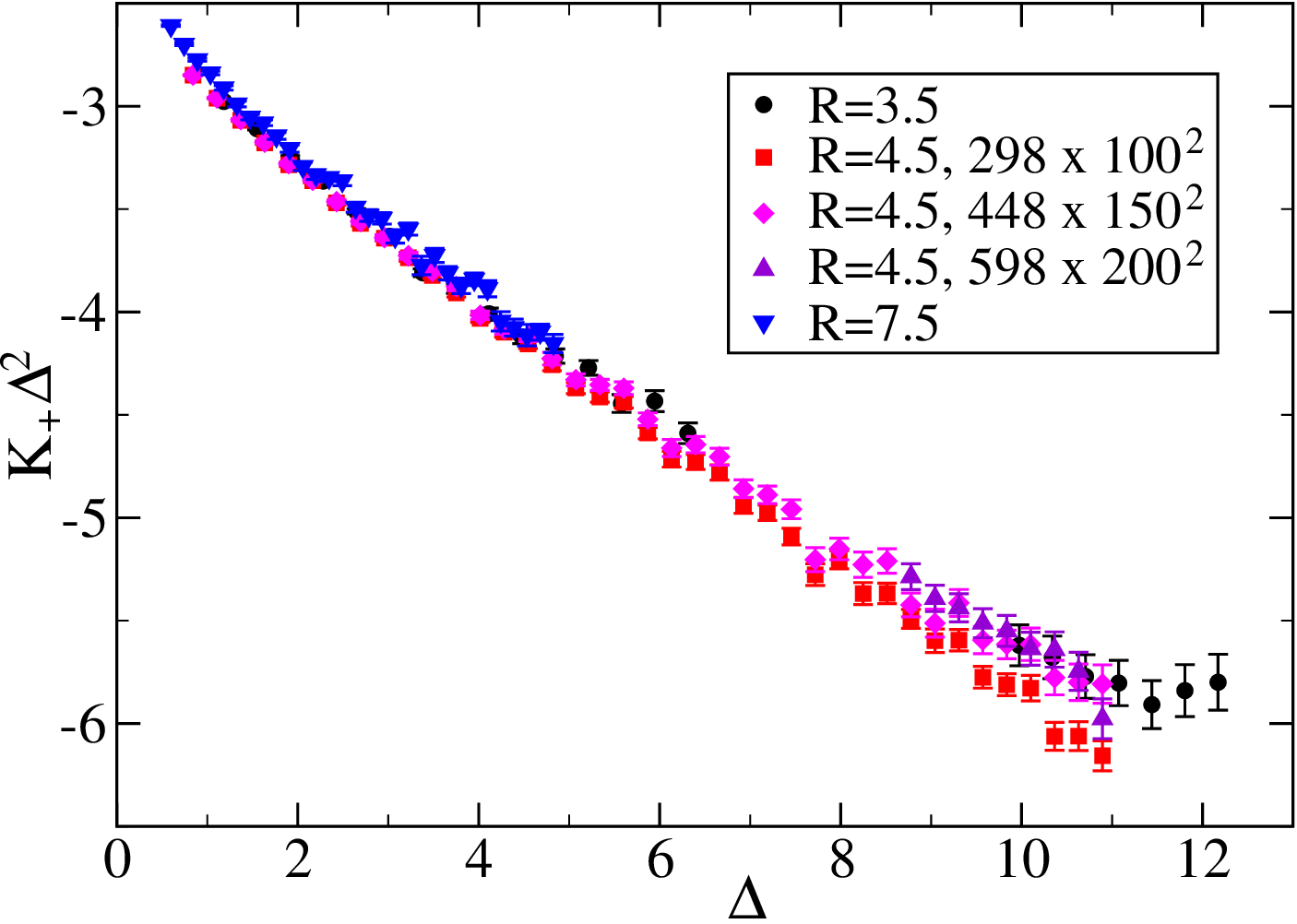}
\caption{\label{bcp}
We plot  $K_+(0,\Delta)  \Delta^2$ as a function of $\Delta$. 
Monte Carlo data obtained by using $R=3.5$, $4.5$, and $7.5$ are given.
In the case of $R=4.5$ we compare results obtained by using the lattice sizes 
$298 \times 100^2$, $448 \times 150^2$, and $598 \times 200^2$. 
Note that $s_{sphere}=1$. 
}
\end{center}
\end{figure}

Finally we discuss the scaling function at the critical point, i.e. $x=0$.  
In figure \ref{bcp} we plot our numerical results for $K_+(0,\Delta) \Delta^2$ as a function of 
$\Delta$. At the critical point we have to consider  finite $L$ and $L_0$ effects. 
Therefore, for $R=4.5$ we plotted our results obtained by using lattices of the 
sizes $298 \times 100^2$, $448 \times 150^2$ and $598 \times 200^2$. In particular for large 
$\Delta$ we see a clear deviation of the results obtained for the $298 \times 100^2$ lattice
from the other two.  On the other hand the estimates obtained for the $448 \times 150^2$ and 
$598 \times 200^2$ lattices are consistent within the statistical error. Therefore deviations 
from the $L_0,L \rightarrow \infty$ limit should be of similar size as the statistical error.
In the case of $R=3.5$ we plot the data for the largest lattice size available.  
We find a quite good collapse of the data obtained for different radii.   For $\Delta=1$ we 
get  $K_+(0,\Delta) \Delta^2 \approx -2.8$. We see that $K_+(0,\Delta) \Delta^2$ is almost
linearly decreasing with increasing $\Delta$. For $\Delta=11$  we read off 
$K_+(0,\Delta) \Delta^2 \approx -6$. The last three data points for $R=3.5$ seem to suggest
that $K_+(0,\Delta) \Delta^2$ is increasing again for $\Delta \gtrapprox 11$. However 
this is not statistically significant and is likely an artifact.

Using the Derjaguin approximation, we get 
$\lim_{\Delta \rightarrow 0}  K_+(0,\Delta)  \Delta^2 \approx -2.59$ and as for $K_{+,min}$, 
$K_+(0,\Delta)  \Delta^2$ is increasing with increasing $\Delta$.  Therefore it seems 
plausible that $K_+(0,\Delta)  \Delta^2$ has a maximum in the range $0 <  \Delta \lessapprox 1$. 

\subsection{The scaling function $K_-(x,\Delta)$ for $s_{sphere}=-1$}
\label{kminus}
\begin{figure}
\begin{center}
\includegraphics[width=14.4cm]{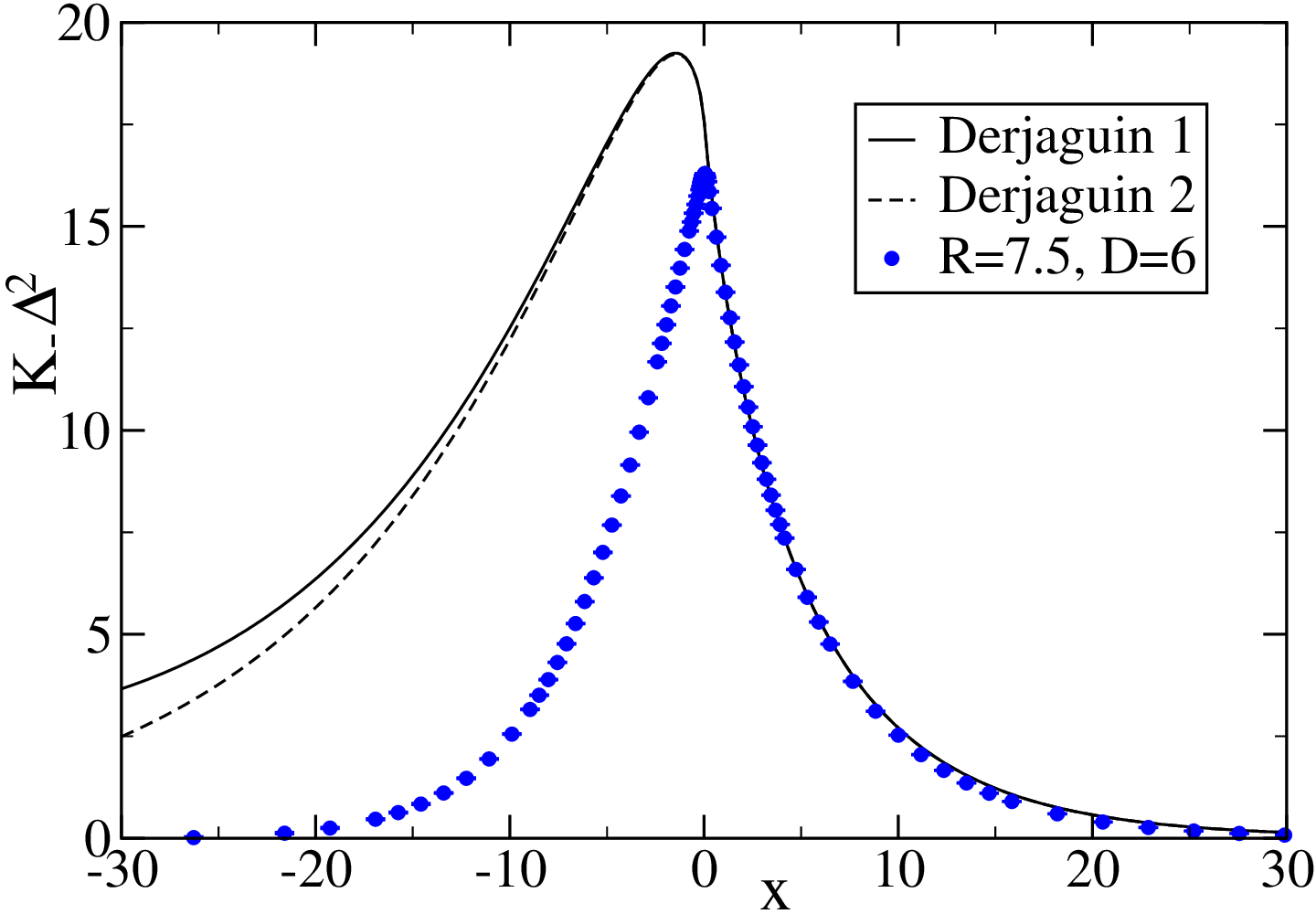}
\caption{\label{Km1}
We plot $K_-(x,\Delta) \Delta^2$  as a function of $x$ for $\Delta \rightarrow 0$
obtained by using the Derjaguin approximation and $\Delta \approx 1$ using our Monte Carlo  
data for $R=7.5$ and $D=6$. Note that the index of $K$ indicates $s_{sphere}=-1$.
In the case of the Derjaguin approximation we have used two different extrapolations of the 
scaling function $\theta_{+-}(x)$  for $x \lessapprox -48$ as discussed in the text.
}
\end{center}
\end{figure}

Next let us turn to the scaling function $K_-(x,\Delta)$ for $s_{sphere}=-1$. 
In figure \ref{Km1} we plot
$K_-(x,\Delta) \Delta^2$ as a function of $x$ for $\Delta \rightarrow 0$
and $\Delta \approx 1$. The limit $\Delta \rightarrow 0$ is obtained by using 
the Derjaguin approximation. Unfortunately we do not have data for the scaling 
function $\theta_{+-}(x)$ for the plate-plate geometry for $x \lessapprox -48$
and the amplitude of $\theta_{+-}(x)$ in this range is still significant. Therefore
we have extrapolated $\theta_{+-}(x)$ in two different ways: (1) $\theta_{+-}(x)=const$
for $x \lessapprox -48$ and (2) $\theta_{+-}(x)$  is linear in the 
interval $-65 < x < -48$ and $\theta_{+-}(-48) = 0.8$, $\theta_{+-}(-65) = 0$, for 
$ x < -65$ we take $\theta_{+-}(x)=0$.  
The true $\theta_{+-}(x)$ should be enclosed by these two extrapolations. In figure 
\ref{Km1}, the results based on these two extrapolations are denoted by 
Derjaguin 1 and Derjaguin 2, respectively.
For comparison we plot our Monte Carlo result obtained for $R=7.5$ and $D=6$ which 
corresponds to $\Delta \approx 1$.  We observe that the two curves are very 
similar in the high temperature phase.  Since $\theta_{+-}(x)$ has its maximum
in the low temperature phase, the same holds for $K_-(x,\Delta)$  computed 
by using the Derjaguin approximation. In contrast, for  $\Delta \approx 1$ we
find that the maximum is located in the high temperature phase, even though very
close to the critical point. Furthermore, $K_-(x,\Delta) \Delta^2$ at 
$\Delta \approx 1$
is much smaller than $\lim_{\Delta \rightarrow 0} K_-(x,\Delta) \Delta^2$
in the low temperature phase. Also the decay of $K_-(x,\Delta) \Delta^2$  
as $x \rightarrow -\infty$ seems to be much faster for $\Delta \approx 1$ than 
for $\Delta \rightarrow 0$.  
\begin{figure}
\begin{center}
\includegraphics[width=14.4cm]{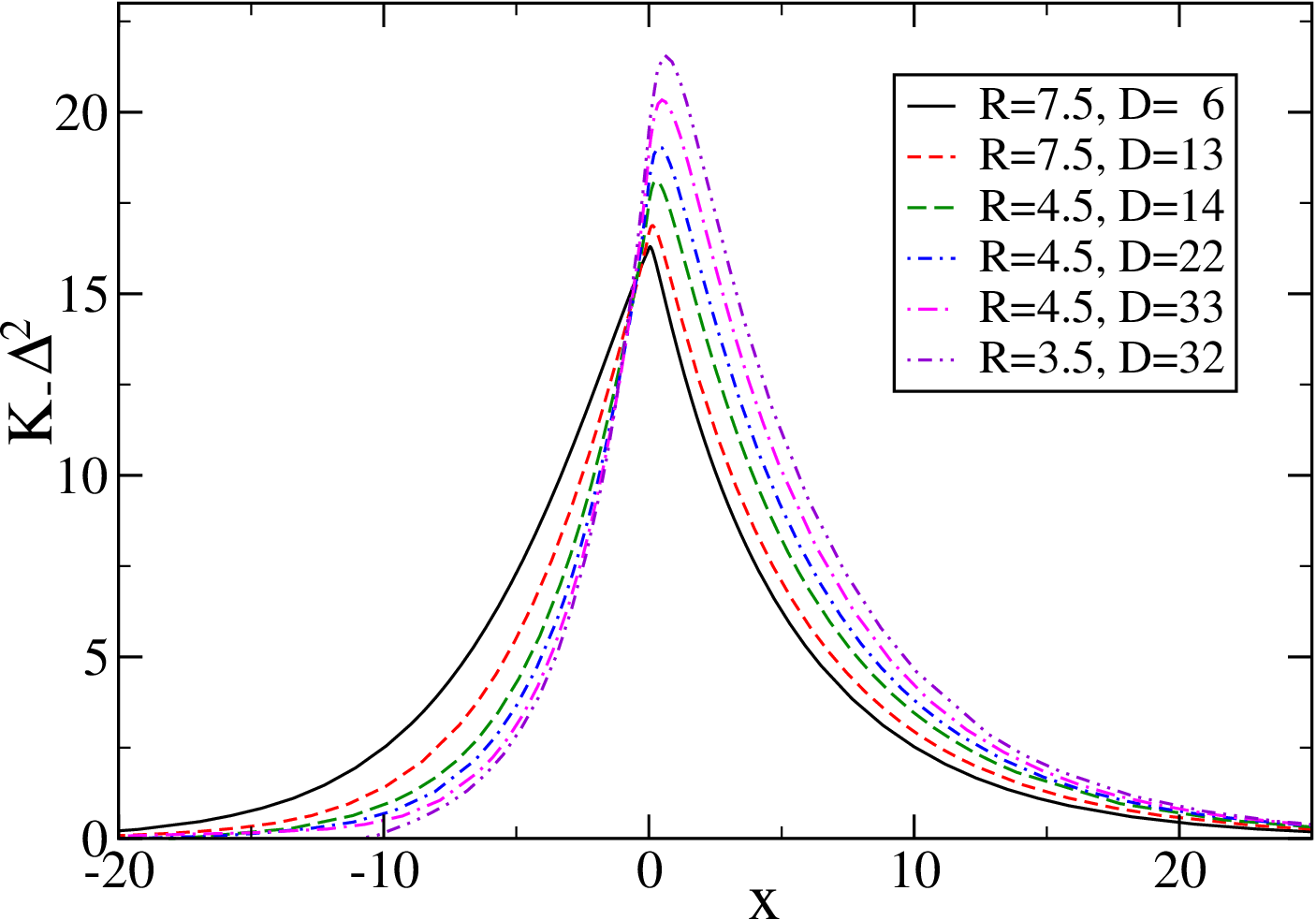}
\caption{\label{Km2}
We plot $K_-(x,\Delta) \Delta^2$  as a function of $x$ for various values of 
$\Delta$. Note that the index of $K$ indicates $s_{sphere}=-1$. We use
Monte Carlo data for $(R,D)$ = $(7.5,6)$, $(7.5,13)$, $(4.5,14)$, $(4.5,22)$,
$(4.5,33)$, and $(3.5,32)$, corresponding to $\Delta \approx$ $1.03$, $2.06$, 
$4.02$, $6.13$, $9.04$, and $12.17$, respectively.
}
\end{center}
\end{figure}

This observation might be explained as follows:
Deep in the low temperature phase the physics of a film with $+-$ boundary conditions
is governed by the interface between the two phases of positive and negative 
magnetisation. In the case of the film, this interface can move quite freely 
between the two plates. In contrast, in the case of the sphere-plate geometry, 
the interface is closely attached to the sphere, minimizing the area of the 
interface.

As for $s_{sphere}=1$ our observations for $s_{sphere}=-1$ are consistent 
with those of ref. \cite{troendle}. For technical reasons the authors of 
\cite{troendle} performed the mean-field calculation only for
$\theta \ge 0$. This means that mean-field results for the low temperature phase, 
where we see a large deviation between the Derjaguin approximation and 
our Monte Carlo result for $\Delta \approx 1$ are
unfortunately missing in Fig. 2 a of \cite{troendle}.

In figure \ref{Km2} we plot
$K_-(x,\Delta) \Delta^2$ as a function of $x$ for various values of $\Delta$.
With increasing $\Delta$ the maximum of $K_-(x,\Delta) \Delta^2$ increases 
and also $x_{max}$ slowly increases. For the range of $\Delta$ studied here, 
$K_-(x,\Delta) \Delta^2$ is increasing with increasing $\Delta$ for $x \gtrapprox -0.6$, while 
it is decreasing  with increasing $\Delta$ for $x \lessapprox -0.6$.

\begin{figure}
\begin{center}
\includegraphics[width=13.5cm]{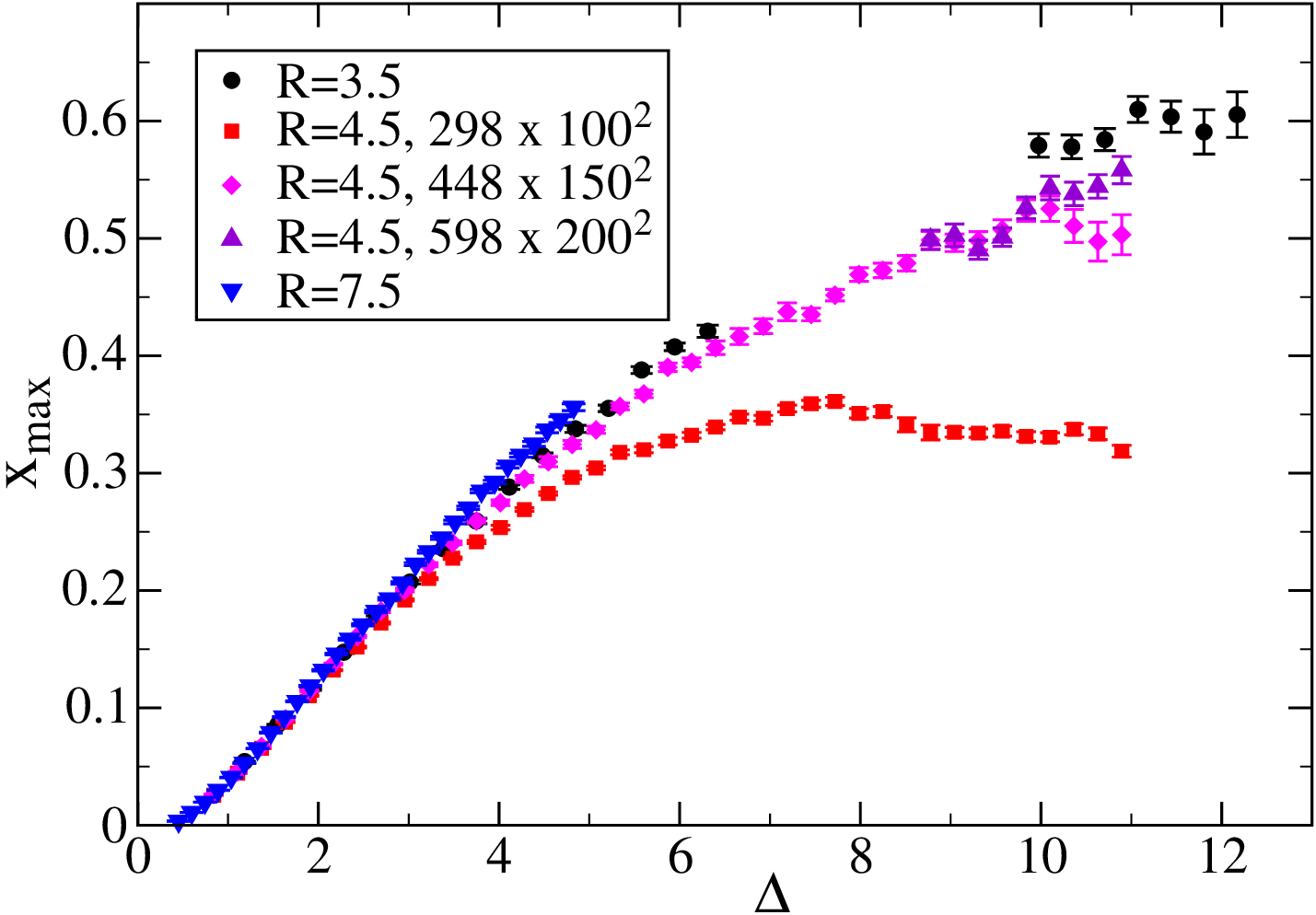}
\caption{\label{xmaxm}
We plot  $x_{max}$ as a function of $\Delta$ for $s_{sphere}=-1$. We give our results
for $R=3.5$, $4.5$ and $7.5$. In the case of $R=4.5$ we compare the results
obtained for the three different lattice sizes
$298 \times 100^2$, $448 \times 150^2$, and $598 \times 200^2$.
}
\end{center}
\end{figure}

Next we study in more detail the maximum of $K_-(x,\Delta)$ as a function of $x$. 
In figure \ref{xmaxm} we plot the location $x_{max}$ of the maximum as a function of $\Delta$.
Since $x_{max}$ is very close to zero, finite $L_0$ and $L$ effects might be large.
In order to check for these effects, we plot for $R=4.5$ the results obtained by using 
lattices of the sizes $298 \times 100^2$, $448 \times 150^2$ and $598 \times 200^2$.
In fact for $\Delta \gtrapprox 3$ we see that the results obtained for the $298 \times 100^2$
lattice clearly deviate from those for the other two lattice sizes. On the other hand, the results
for the lattice sizes $448 \times 150^2$ and $598 \times 200^2$ are essentially consistent and 
are more or less consistent with those for $R=3.5$. Therefore we consider the results that we 
have obtained for our largest lattice sizes as reasonable approximation of the 
$L_0,L \rightarrow \infty$ limit.  

Using the Derjaguin approximation, we get $x_{max} \approx -1.43 $ in the limit
$\Delta \rightarrow 0$. Within the Derjaguin approximation $x_{max}$ is decreasing 
with increasing $\Delta$. In contrast for all values of $\Delta$ that we studied
by Monte Carlo simulations, we find $x_{max} > 0$.  Looking at our data obtained
for $R=7.5$ we estimate $x_{max} = 0$ at $\Delta \approx 0.5$. $x_{max}$ slowly 
increases with increasing $\Delta$. For $\Delta \approx 12$ we get $x_{max} \approx 0.6$.

\begin{figure}
\begin{center}
\includegraphics[width=13.5cm]{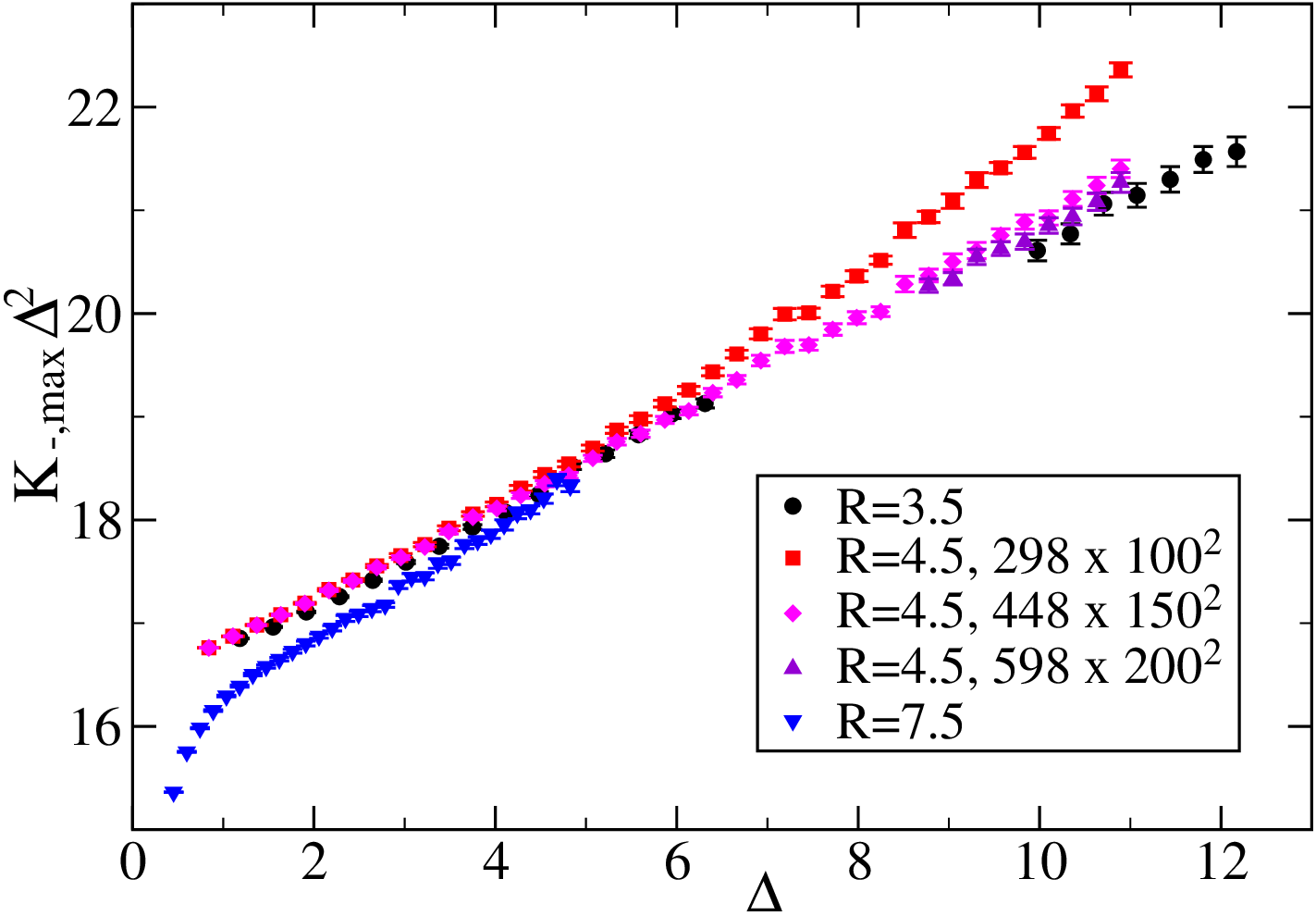}
\caption{\label{maxm}
We plot $K_-(x_{max},\Delta)  \Delta^2$ as a function of $\Delta$ for $s_{sphere}=-1$.
We give our results
for $R=3.5$, $4.5$ and $7.5$. In the case of $R=4.5$ we compare the results
obtained for the three different lattice sizes $298 \times 100^2$, $448 \times 150^2$, 
and $598 \times 200^2$.
}
\end{center}
\end{figure}
\begin{figure}
\begin{center}
\includegraphics[width=13.5cm]{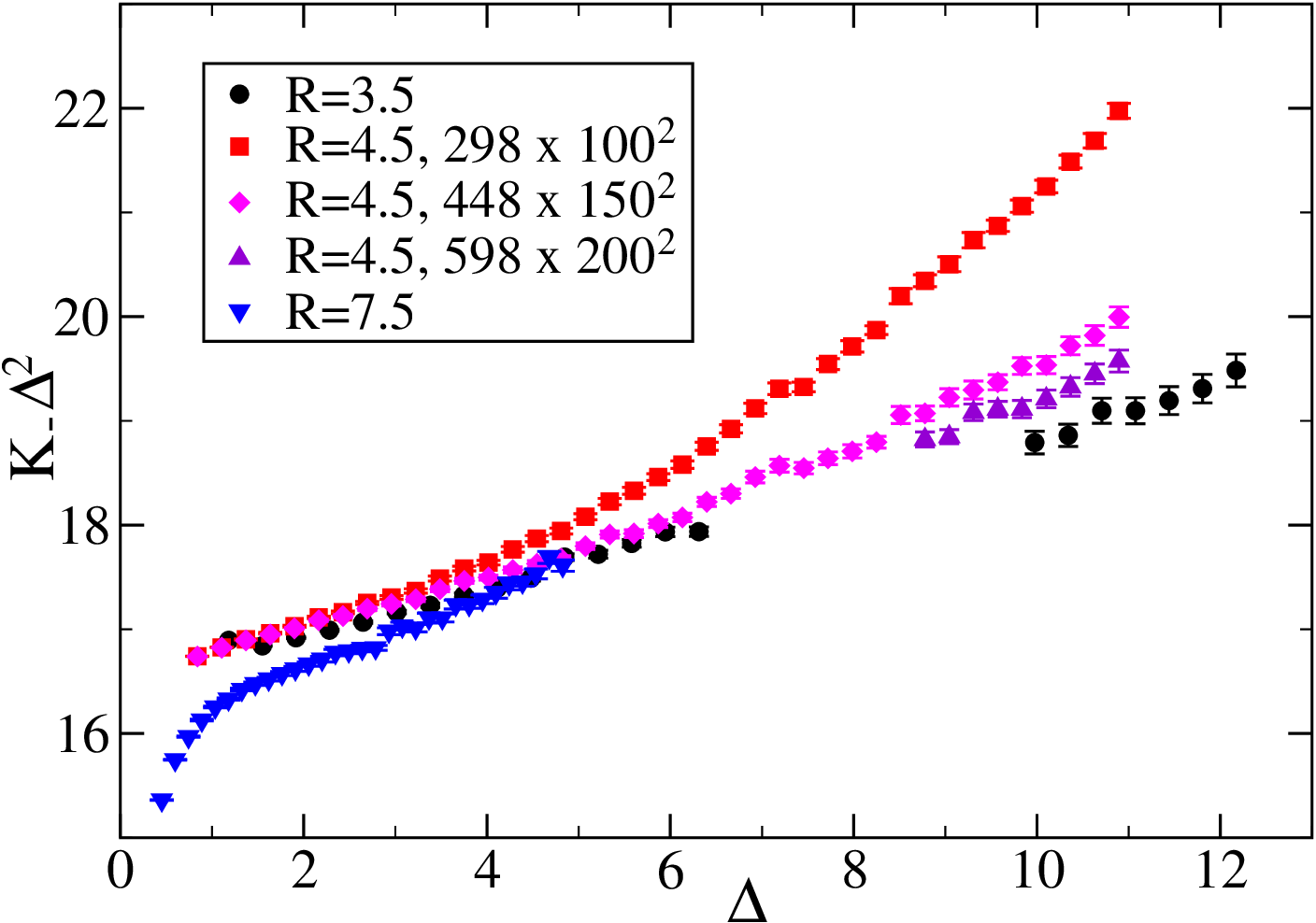}
\caption{\label{bcm}
We plot $K_-(0,\Delta) \Delta^2$ as a function of $\Delta$ for $s_{sphere}=-1$.
We give our results
for  $R=3.5$, $4.5$ and $7.5$. In the case of $R=4.5$ we compare the results
obtained for the three different lattice sizes $298 \times 100^2$, $448 \times 150^2$,
and $598 \times 200^2$.
}
\end{center}
\end{figure}

In figure \ref{maxm} we plot 
$K_-(x_{max},\Delta) \Delta^2$ as a function of $\Delta$.  Also here we check for finite
$L_0$ and $L$ effects by plotting for $R=4.5$ the results obtained for three different 
lattice sizes. Also here we find that the result for the $298 \times 100^2$ lattice 
differs from those for the $448 \times 150^2$ and $598 \times 200^2$ lattices for larger 
values of $\Delta$, while the results obtained for lattices of the sizes $448 \times 150^2$ and 
$598 \times 200^2$ are in quite good agreement.  For $\Delta \lessapprox 4$ the results 
for $R=7.5$ and the other radii do not agree within error bars indicating violations of scaling 
that are larger than the numerical errors. Taking into account these observations we conclude 
that $K_-(x_{max},\Delta) \Delta^2$ increases roughly linearly in the range of $\Delta$ that we
studied. We get $K_-(x_{max},\Delta) \Delta^2 \approx 16$  for $\Delta \approx 1$ and 
$K_-(x_{max},\Delta) \Delta^2 \approx 21.5$ for $\Delta \approx 12$.
In the limit $\Delta \rightarrow 0$ we get  $K_-(x_{max},\Delta) \Delta^2  \approx 19.2$.
Within the Derjaguin approximation $K_-(x_{max},\Delta) \Delta^2$ is
decreasing with increasing $\Delta$. Hence $K_-(x_{max},\Delta) \Delta^2$ should reach a minimum
in the interval $0 < \Delta \lessapprox 1$.

In figure \ref{bcm} we plot
$K_-(0,\Delta) \Delta^2$ as a function of $\Delta$.
Our observations concerning finite $L_0$ and $L$ effects and scaling are qualitatively similar
as for $K_-(x_{max},\Delta)  \Delta^2$. However the amplitudes of the differences are larger than 
for  $K_-(x_{max},\Delta)  \Delta^2$.  For large values of $\Delta$ even the results for 
the lattice sizes $448 \times 150^2$ and $598 \times 200^2$ do not agree within the statistical
error.  Taking into account these observation we conclude that $K_-(0,\Delta)  \Delta^2$
is increasing roughly linearly in the range $1 \lessapprox \Delta \lessapprox 12$. We get
$K_-(x_{max},\Delta)  \Delta^2 \approx 16$ for $\Delta \approx 1$ and 
$K_-(x_{max},\Delta)  \Delta^2 \approx 19.5$ for $\Delta \approx 12$.   In the Derjaguin 
approximation we get $\lim_{\Delta \rightarrow 0} K_-(0,\Delta)  \Delta^2 \approx 17.6$.
In the Derjaguin approximation, $K_-(0,\Delta)  \Delta^2$ is decreasing with 
increasing $\Delta$.  Therefore, as it is the case for $K_-(x_{max},\Delta)  \Delta^2$, 
$K_-(0,\Delta)  \Delta^2$ should show a minimum in the interval $0 < \Delta \lessapprox 1$. 

\subsection{Small sphere expansion at the critical point}
\label{SecSmall}

In the limit $D \gg R$ the behaviour of the reduced excess free energy at the critical point 
is given by \cite{BuEi97}
\begin{equation}
\label{excessfree}
 F_{ex}(D,R) = - \frac{A_a^{\psi} A_b^{\psi}}{B_{\psi}} \left(\frac{R}{2 D} \right)^{x_{\psi}}
\end{equation}
where $x_{\psi}$ is the dimension of the field $\psi$. The amplitudes are defined by
\begin{equation}
\label{amplitude1}
 \langle \psi_0 \psi_r \rangle_{bulk}  = B_{\psi} r^{-2 x_{\psi}}
\end{equation}
and 
\begin{equation}
\label{amplitude2}
 \langle  \psi_r \rangle_{half space}^a  = A_a^{\psi} (2 r)^{-x_{\psi}}
\end{equation}
where in the case of eq.~(\ref{amplitude2}) $r$ is the distance from 
the boundary of type $a$. For the boundary conditions discussed 
here, the leading contribution comes from the order parameter $\phi$ with 
\begin{equation}
x_{\phi} = \beta/\nu=\frac{1}{2} (d - 2 +\eta)
\end{equation}
where $\eta=0.03627(10)$ \cite{mycritical}. The subleading contribution 
is due to the energy density $\epsilon$ with 
\begin{equation}
x_{\epsilon} = d - 1/\nu
\end{equation}
where $\nu=0.63002(10)$ \cite{mycritical}.

In relation with ref. \cite{mycrossover} we had simulated $512^3$ lattices
with $++$ and $+0$ boundary conditions in one direction and periodic ones
in the other two directions at $\beta=0.387721735$. 
We had measured the magnetisation profile.
Taking into account the extrapolation length, we get from the magnetisation
profile at $r \approx 30$ the estimate
\begin{equation}
\label{Aphi}
A_+^{\phi} = 1.1268(5)  \;\;.
\end{equation}
In order to compute the amplitude $B_{\phi}$ we simulated systems
with periodic boundary conditions in all three directions.  We  simulated
lattices of the sizes $128^3$, $256^3$, $512^3$ and $1024^3$ 
at $\beta=0.387721735$. We measured $\langle s_x s_y \rangle$ for
$y-x =(i,0,0)$, $(0,i,0)$ and $(0,0,i)$ and $i=1, 2, ...$ . Using the numerical
results of the two-point function we computed
\begin{equation}
B_{\phi,eff}(i) = i^{2 x_\phi} \langle s_x s_{x+(i,0,0)} \rangle 
\end{equation}
using $2 x_\phi = 1.03627(10)$. It turns out that $B_{\phi,eff}(i)$  shows
a shallow minimum for a rather small value of $i$. At larger distances, 
$B_{\phi,eff}(i)$ increases due to the periodic boundary conditions imposed.
The minimum is given by $(i_{min},B_{\phi,eff,min}) = $
$(4, 0.19397(8) )$,
$(6, 0.19026(15) )$,
$(7, 0.18838(11) )$, and    
$(8, 0.18745(13) )$ for lattices of the size $128^3$, $256^3$, $512^3$ and $1024^3$,
respectively. Assuming  power like corrections, we arrive at our final estimate
\begin{equation}
\label{Bphi}
 B_{\phi}  = 0.1865(3) \;\;.
\end{equation}
Combining eqs.~(\ref{Aphi},\ref{Bphi}) we arrive at 
\begin{equation}
\label{supera}
a = \frac{A_+^{\psi} A_+^{\psi}}{B_{\psi}} = 6.808(12) \;\;.
\end{equation}
This result can be  compared with $a \approx 7.73$ given in ref. \cite{HaScEiDi98}, where the authors 
had deduced
the estimate for three dimensions from the exact result in two dimensions and the 
$\epsilon$-expansion. For a discussion see footnote [13] of ref. \cite{HaScEiDi98}.

As check we computed $Z_{-}/Z_{+}$ for the radii $R=3.5$ 
using the sphere flip algorithm discussed in Appendix \ref{App2}.  We have
simulated lattices of the size $L_0 \times L^2$ where we have fixed $s_x=1$ for $x_0=0$ and 
$s_x=0$ for $x_0=L_0+1$. Our numerical results for $R=3.5$ are summarized in table \ref{Z+-/Z++}.
First we have simulated for $h=D+R=72$ lattices of different linear sizes $L_0$ and $L$ in order
to check numerically the effect of the finite size of the lattice. We expect that at the critical 
point, the result for $Z_{-}/Z_{+}$ converges with a power law to the $L_0, L \rightarrow \infty$ 
limit. The results of $(L_0, L)=(499,400)$ and $(L_0, L)=(799,600)$ are consistent within the 
statistical error.  Next we simulated lattices of the size $(L_0, L)=(499,400)$ for various values 
of $h=D+R$.

Inserting $A_+^{\psi} = - A_-^{\psi}$ into eq.~(\ref{excessfree}) we get for $D \gg R$
\begin{equation}
\label{excessfree2}
- \ln(Z_{-}/Z_{+}) = 2 \frac{A_+^{\psi} A_+^{\psi}}{B_{\psi}} \left(\frac{R}{2 D} \right)^{x_{\psi}}
\;\;.
\end{equation}
For finite $D$ we define
\begin{equation}
\label{aD}
 a(D) = - \frac{1}{\ln(Z_{-}/Z_{+})} \left(\frac{2 D_{eff}}{R_{eff}} \right)^{x_{\psi}} \;\;,
\end{equation}
where we replaced $R$ and $D$ by $R_{eff}$, table \ref{refftab2}, and $D_{eff}$, eq.~(\ref{Deff}), 
respectively, to reduce corrections to scaling.
\begin{table}
\caption{\sl \label{Z+-/Z++}  Results for the ratio $Z_{-}/Z_{+}$ of partition functions for the 
radius $R=3.5$. In the fourth column we give the number of measurements. In the last column we give
the estimate of the universal ratio $a$. In $()$ we give the error due to the statistical error of 
$Z_{-}/Z_{+}$ and in $[]$ the error due to the uncertainty of $R_{eff}$.  For a discussion see the 
text.
}
\begin{center}
\begin{tabular}{ccccll}
\hline
$h=D+R$ & $L_0$&  $L$  & stat& $Z_{-}/Z_{+}$ & \phantom{aaa} $a$ \\
\hline
 72 & 199 & 160   & $10^5$ & 0.1645(28)       & \\ 
 72 & 299 & 250   & $10^5$ & 0.1472(26)       & \\ 
 72 & 359 & 300   & $10^5$ & 0.1449(26)       & \\
 72 & 499 & 400   & $1.5 \times 10^5$ & 0.1391(21) & 7.57(6)[9] \\ 
 72 & 799 & 600   & $10^5$ & 0.1384(25)       &  \\ 
\hline
 30 & 499 & 400   & $1.2 \times 10^5$      &0.0289(7) & 8.43(6)[11] \\ 
 40 & 499 & 400   & $1.2 \times 10^5$      &0.0552(12) & 8.09(6)[10] \\ 
 50 & 499 & 400   & $1.2 \times 10^5$      &0.0761(15) & 8.12(6)[10] \\ 
 60 & 499 & 400   &    $10^5$    & 0.1090(20)          & 7.71(6)[9] \\ 
 90 & 499 & 400   & $1.2 \times 10^5$        &0.1830(27) & 7.34(6)[9] \\ 
\hline
\end{tabular}
\end{center}
\end{table}

In the last column of table \ref{Z+-/Z++} we give estimates of $a$ obtained 
for various distances $D$. It is decreasing with increasing distances. For 
the largest distance that we simulated, it is still clearly larger than the 
estimate~(\ref{supera}), indicating that still at $\Delta \approx 30$ subleading
contributions to $F_{ex}$ are significant. 

\section{Conclusions and outlook}
\label{conclus}
We have demonstrated how the thermodynamic Casimir force for the sphere-plate 
geometry can be computed efficiently by Monte Carlo simulations of spin models.
Similar to the proposal of Hucht \cite{Hucht} for the plate-plate geometry, 
we compute differences of the free energy by integrating differences of the energy
over the inverse temperature. For $R \ll L_0,L$, where $R$ is the radius of the sphere and 
$L_0,L$ are the linear extensions of the lattice, such energy differences, 
when determined the standard way, are affected by a huge relative variance. Here we demonstrate
how this problem can be overcome by using the exchange cluster algorithm \cite{HeBl98}. 
Using the exchange cluster algorithm we define an improved estimator for the 
energy differences with a strongly reduced variance. For a detailed discussion see 
section \ref{algorithm}. 
We  simulated the improved Blume-Capel model, which shares the universality class of the 
three-dimensional Ising model. Improved means that the amplitudes of the leading 
bulk correction are strongly suppressed.
We studied strongly symmetry breaking boundary 
conditions. We fixed the spins at the boundary to $s_x=1$. The spins at the 
surface of the sphere are fixed to $s_x = s_{sphere}$, where we studied the two
cases $s_{sphere}=-1$ and $1$ in detail. First we verified that  the
method is efficient for a large range of parameters. We studied the whole
range of temperatures where the thermodynamic Casimir force has a significant 
amplitude. We considered distances between the sphere and the plate up to $D_{max} = 32$, 
$40$, $32$ and $10$ for the radii  $R=3.5$, $4.5$, $7.5$ and $15.5$, respectively.
In order to see a good scaling behaviour of the data obtained for different 
radii, we introduced an effective radius $R_{eff}$ of the spheres that 
we determined by using a particular finite size scaling analysis. For details 
see  Appendix \ref{App1}.  We obtain reliable results for the scaling 
functions $K_+(x,\Delta)$ and $K_-(x,\Delta)$ for the whole range of $x$ 
and $1 \lessapprox \Delta \lessapprox 12$, where $x=t [D/\xi_0]^{1/\nu}$ 
and $\Delta =D/R$.  Smaller values of $\Delta$ would be desirable to make 
better contact with the Derjaguin approximation and larger ones to reach 
the range of validity of the small sphere expansion. For $s_{sphere}=1$ 
we find that for the values of $\Delta$ studied here, 
the shape of $K_+(x,\Delta) \Delta^2$ viewed as a function of $x$ is quite 
similar to that of $\lim_{\Delta \rightarrow 0} K_+(x,\Delta) \Delta^2$ 
obtained by using the Derjaguin approximation.  For $s_{sphere}=-1$
in the high temperature phase still the same observation holds, however 
in the low temperature phase  $K_-(x,\Delta) \Delta^2$ 
is strongly suppressed  for $\Delta \gtrapprox 1$ compared with 
$\Delta \rightarrow 0$.  We attribute this behaviour to the fact that 
the interface between positive and negative magnetisation is closely
attached to the sphere, in order to minimize the area of the interface.
In contrast, for the film geometry, the interface is moving quite 
freely between the boundaries.
Finally, by using a different cluster algorithm, which is similar to 
the boundary flip algorithm \cite{MH93A,MH93B}, we computed the difference 
of reduced free energies $F_--F_+$ for large distances $D$ at the critical 
point, where the subscript of $F$ gives the sign of $s_{sphere}$.  Here 
we made contact with the small sphere expansion. Still for $\Delta \approx 30$
we see a deviation from the leading order of the small sphere expansion 
at the level of about $7 \%$.

The present work should be seen as a pilot study. It can be improved and 
extended in many respects. 
Our preliminary study for $s_{sphere}=0$ discussed in section \ref{clustersize}
indicates that for $s_{sphere}=0$ the performance of the exchange 
cluster algorithm is worse than for $s_{sphere}=\pm 1$, however still it
seems likely that
physically relevant results can be obtained. Based on this experience 
we are confident that the algorithm discussed here can be applied to a variety of
experimentally relevant situations. E.g. the crossover of boundary universality
classes or patterned surfaces. In particular also the sphere-sphere geometry 
should be accessible by the algorithm discussed here. One might also
check whether the exchange cluster algorithm allows to improve the results
for the film geometry that have been obtained so far using standard simulation
algorithms.

One could improve the present study by various means. In order to reach larger 
lattice sizes one might implement the computer program by using the multispin-coding 
technique. To this end one might also try to simulate with a varying lattice
resolution: spins far away from the sphere could be replaced by block-spins.
In order to reduce corrections to scaling one might try to improve the spherical
shape of the sphere by tuning the couplings at the surface of the sphere.

\section{Acknowledgement}
This work was supported by the DFG under the grant No HA 3150/2-2.

\appendix

\section{Determination of the effective radius $R_{eff}$}
\label{App1}
In order to determine the effective radius of the sphere, we  studied
the ratio of partition functions $r=Z_{-}/Z_{+}$ for a lattice 
of the linear sizes $L_0$ and $L_1=L_2=L$, where the sphere is placed exactly
in the middle of the system; i.e. $h=(L_0+1)/2$ for odd values of $L_0$. The spins at $x_0=0$ and 
$x_0=L_0+1$ are fixed to $s_x=1$. The  index of $Z$ refers to the sign of $s_{sphere}$.
At the critical point, in the scaling limit
\begin{equation}
\label{scalingr}
 r(R,L_0,L)= G(R/L_0,L/L_0) \;\;.
\end{equation}
In the following, we restrict ourselves to $L/L_0=1/2$ and define $g(R/L_0) = G(R/L_0,1/2)$. 
Furthermore we require that the ratio of partition functions assumes a fixed value. Our ad hoc choice 
is $r=0.1$. For a given radius $R$ this fixes the linear size of the lattice to $L_0=L_{0,fix}$. 
In order to reduce corrections to 
scaling we replace $L_0$ by $L_{0,eff} = L_0+L_s$ using $L_s=1.92$ \cite{mycrossover}.
Since $L$ assumes integer and $L_0$ odd integer values, we interpolate or extrapolate $r(R,L_0,L)$ to
$r(R,L_0,L_{0,eff}/2)$.
To this end, we simulated for $R=3.5$ and $L_0=131$ the linear extensions
$L=64$, $66$ and $68$. The results are $r=0.10405(23)$, $0.10015(16)$ and
$0.09641(22)$, respectively. We performed $10^7$, $2 \times 10^7$ and $10^7$
measurements for $L=64$, $66$ and $68$, respectively. In order to apply the
result also to other radii, we compute
\begin{equation}
\label{extra1}
\left . \frac{\partial r}{\partial (L/L_{0,eff})} \right |_{L/L_{0,eff}=0.5,L_0=L_{0,fix}}
 = - 0.254(11)
\end{equation}
where we have used the finite difference of the estimates for $L=64$ and $68$. For
larger radii we have simulated $L=(L_0+1)/2$ and have extrapolated $r$ to $L=(L_0+1.92)/2$
by using eq.~(\ref{extra1}). 
In addition to  $R=3.5$ we have simulated
the radii $R=4.5$, $5.5$, $6.5$, $7.5$, $9.5$, $12.5$ and $15.5$. We performed
$8 \times 10^6$ measurements for $R=4.5$ down to $3.4 \times 10^6$ measurements
for $R=15.5$. In total these simulations took about 5 years of CPU-time
on a single core of a Quad-Core AMD Opteron(tm) 2378 CPU running at 2.4 GHz. Our results for 
$L_0$ such that $r \approx 0.1$ are summarized in table  \ref{rrrr}.  
We linearly extrapolate $r(R,L_0,L_{0,eff}/2)$ in $L_0$. To this end, we use
\begin{equation}
L_{0,eff} \frac{\partial r(R,L_0,L_{0,eff}/2)}{\partial L_0} = 0.11(1)
\end{equation}
which we have computed by using finite differences. Solving $r(R,L_0,L_{0,eff}/2)=0.1$ with respect
to $L_0$ we arrive at the estimate of $L_{0,eff,fix}$ given in the last column of table \ref{rrrr}.

\begin{table}
\caption{\sl \label{rrrr}   Estimates of $r(R,L_0,L_{0,eff}/2)$  for $L_0$ such that
$r \approx 0.1$.  In the last column we give the solution of $r(R,L_0,L_{0,eff}/2)=0.1$.
For a discussion see the text.
}
\begin{center}
\begin{tabular}{rccl}
\hline
 $R$\phantom{a}  & $L_0$ &      $r$    & $L_{0,eff,fix}$    \\
\hline
 3.5 &  131 &   0.09927(20) &     133.8(3)      \\
 4.5 &  183 &   0.09972(30) &     185.4(6)      \\
 5.5 &  235 &   0.09922(32) &     238.6(0.8)    \\
 6.5 &  287 &   0.10162(32) &     284.7(1.3)    \\
 7.5 &  331 &   0.09903(39) &     335.9(1.5)    \\
 9.5 &  435 &   0.10027(41) &     435.8(1.7)    \\
11.5 &  535 &   0.10103(43) &     531.9(2.6)    \\
15.5 &  731 &   0.10065(57) &     728.6(4.2)    \\
\hline
\end{tabular}
\end{center}
\end{table}

Now we define the effective radius by 
\begin{equation}
\label{Reffeq}
 R_{eff} = c^{-1}  L_{0,eff,fix}(R) 
\end{equation}
In order to determine the factor $c$ we fitted our data with the Ansatz
\begin{equation}
\label{reff1}
 L_{0,eff,fix}(R) = c [R + R_s] \;\;.
\end{equation}
where $c$ and $R_s$ are the parameters of the fit and with
\begin{equation}
\label{reff2}
 L_{0,eff,fix}(R) = c [R + R_s] \times (1 + b [R + R_s]^{-2})
\end{equation}
with the additional free parameter $b$. The results of our fits are summarized in 
table \ref{refftab}. As our final estimate we take $c=49(1)$ which is consistent 
with all results given in table \ref{refftab}.
\begin{table}
\caption{\sl \label{refftab} Results of fits with the Ans\"atze~(\ref{reff1},\ref{reff2}). 
All data with $R_{min} \le R \le 15.5$ are taken into account. In the case of the 
Ansatz~(\ref{reff2}) we do not report the estimates of $b$. 
}
\begin{center}
\begin{tabular}{ccccc}
\hline
Ansatz & $R_{min}$  &  $c$   &  $R_s$   & $\chi^2/$d.o.f. \\
\hline
\ref{reff1} & 4.5  &   49.65(22) & --0.74(2) &  3.11  \\
\ref{reff1} & 5.5  &   49.08(28) & --0.65(4) &  1.19  \\
\ref{reff1} & 6.5  &   49.47(37) & --0.72(6) &  0.78  \\
\hline
\ref{reff2} & 3.5  &   49.00(31) & --0.61(5) &  1.97  \\
\ref{reff2} & 4.5  &   48.57(41) & --0.52(8) &  2.07  \\
\hline
\end{tabular}
\end{center}
\end{table}
Finally in table \ref{refftab2} we summarize the results for the effective radii $R_{eff}$
for the values of $R$ that are studied in the remainder of the paper.

\begin{table}
\caption{\sl \label{refftab2}  Results for the effective radius $R_{eff}$ obtained by using 
eq.~(\ref{Reffeq}) and the estimate $c=49(1)$.
}
\begin{center}
\begin{tabular}{ccccc}
\hline
  $R$   &  $R_{eff}$ \\
\hline
  3.5   &  2.73(6)   \\
  4.5   &  3.78(9)   \\
  7.5   &  6.86(17)  \\
 15.5   & 14.87(38)  \\
\hline
\end{tabular}
\end{center}
\end{table}

\section{The sphere flip algorithm}
\label{App2}
In Appendix \ref{App1} and section \ref{SecSmall} we have computed the ratio 
\begin{equation}
 r= \frac{Z_{-}}{Z_{+}} 
\end{equation}
by using a special cluster algorithm. The  index of the partition function
$Z$ refers to the sign 
$s_{sphere}$ of the sphere. This cluster algorithm is closely related to the 
boundary flip algorithm discussed in refs. \cite{MH93A,MH93B} that allows to 
compute the ratio $Z_a/Z_p$, where $p$ refers to periodic and $a$ to anti-periodic
boundary conditions. In ref. \cite{mycorrection} we had employed a similar algorithm
to compute $Z_{-+}/Z_{++}$, where the indices of $Z$ refer to the 
signs of the spins at the boundaries.

Let us consider the composed system 
\begin{equation}
 Z = Z_{-} + Z_{+} = \sum_{s_{sphere} = \pm 1} \sum_{\{s\}} \exp(-H(\{s\},s_{sphere}))
\end{equation}
where $s_{sphere}$ can be viewed as an Ising spin. This system can be updated by using the 
standard cluster algorithm. When the cluster that contains the sphere is not frozen 
to the boundary, $s_{sphere}$ can be flipped. In this composed system, the ratio
$r=Z_{-}/Z_{+}$ is given by the ratio of expectation values 
$r=\langle \delta_{-1,s_{sphere}} \rangle /\langle \delta_{1,s_{sphere}} \rangle$. 

In fact it is sufficient to simulate $Z_{+}$ only. Since for 
$s_{sphere}=-1$ the cluster that contains the sphere never freezes to the boundary it 
is sufficient to compute the fraction of cluster updates that would allow to update 
$s_{sphere}=1$ to $s_{sphere}=-1$. To this end, on has to check for each cluster 
update, whether the cluster that contains the sphere is frozen to the boundary. 
In the composed system 
\begin{equation}
\frac{Z_{-}}{Z_{+}} = \frac{p(+ \rightarrow -)}{p(- \rightarrow +)} = p(+ \rightarrow -)
\end{equation}
where $p(- \rightarrow +)$ and $p(- \rightarrow +)$ are the probabilities to update
$s_{sphere}=-1$ to $s_{sphere}=1$ and vice versa, respectively. As discussed above
$p(- \rightarrow +)=1$. 

We have updated the $s_{sphere}=1$ system by a combination of the local heat-bath algorithm and 
the Swendsen-Wang 
cluster algorithm \cite{SwWa87}. In the case of the Swendsen-Wang cluster algorithm \cite{SwWa87} the 
boundary conditions have to be taken into account:
All spins that belong to clusters that are frozen to the boundary or 
contain the sphere keep their value. All other clusters are flipped with the probability $1/2$,
where flipped means that the sign of all spins in a given cluster is changed.

\end{document}